\documentclass[MSNbibl,nameyear,seceqn,dvips]{arxbj}
\usepackage{graphicx}
\aid{0}
\volume{19}
\issue{4}
\pubyear{2013}
\firstpage{1212}
\lastpage{1242}
\doi{10.3150/12-BEJSP11} 

\makeatletter

\define@key{bid}{pmid}{}

\renewcommand{\citep}[1]{(\citeauthor{#1}, \citeyear{#1})}
\newcommand{\citepp}[1]{\citeauthor{#1} \citeyear{#1}}

\newremark{rem}{Remark}[section]

\newcommand{\PP}{\mathbb{P}}
\newcommand{\EE}{\mathbb{E}}
\newcommand{\R}{\mathbb{R}}
\newcommand{\Nat}{\mathbb{N}}
\newcommand{\eps}{\varepsilon}
\newcommand{\Var}{\operatorname{Var}}
\newcommand{\by}{\mathbf{Y}}
\newcommand{\bx}{\mathbf{X}}
\newcommand{\argmin}{\operatorname{argmin}}

\newtheorem{theo}{Theorem}
\newtheorem{prop}{Proposition}
\newtheorem{lemm}{Lemma}
\newtheorem{corr}{Corollary}

\newproclaim{defi}{Definition}
\newremark{examp}{Example}
\makeatother

\begin{document}
\begin{frontmatter}

\title{Statistical significance in high-dimensional linear models}
\runtitle{Significance in high-dimensional models}

\begin{aug}
\author{\fnms{Peter} \snm{B\"uhlmann}\corref{}\ead[label=e1]{buhlmann@stat.math.ethz.ch}}
\runauthor{P. B\"uhlmann} 
\address{Seminar f\"ur Statistik, HG G17, ETH Z\"urich, CH-8092
Z\"urich, Switzerland.\\
\printead{e1}}
\end{aug}


%
\begin{abstract}
We propose a method for constructing $p$-values for general hypotheses in a
high-dimensional linear model. The hypotheses can be local for testing a
single regression parameter or they may be more global involving
several up
to all parameters. Furthermore, when considering many hypotheses, we
show how to adjust for multiple testing taking dependence among the
$p$-values into account. Our technique is based
on Ridge estimation with an additional
correction term due to a substantial projection bias in high
dimensions. We
prove strong error control for our $p$-values and provide sufficient
conditions for detection:
for the former, we
do not make any assumption on the size of the true underlying regression
coefficients while regarding the latter, our procedure might not be
optimal in
terms of power. We demonstrate the method in simulated examples and a real
data application.
\end{abstract}

%

\end{frontmatter}

\section{Introduction}

Many data problems nowadays carry the structure that the number $p$ of
covariables may greatly exceed sample size $n$, i.e., $p \gg n$. In
such a
setting, a huge amount of work has been
pursued addressing prediction of a new response variable, estimation of an
underlying parameter vector and variable selection, see for example the
books by \citet{hastetal09}, \citet{pbvdg11} or the more specific review
article by \citet{fanlv10}. With a few exceptions, see
Section~\ref{subsec.otherwork}, the proposed methods and presented mathematical
theory do not address the problem of assigning uncertainties, statistical
significance or confidence: thus, the area of statistical hypothesis
testing and construction of confidence intervals is largely unexplored and
underdeveloped. Yet, such significance or confidence measures are
crucial in applications where interpretation of parameters and
variables is very important. The focus of this paper is the
construction of
$p$-values and corresponding multiple testing adjustment for a
high-dimensional linear model which is often very useful in $p \gg n$
settings:
%
\begin{equation}
\label{mod.lin} \by= \bx\beta^0 + \eps,
\end{equation}
where $\by= (Y_1,\ldots,Y_n)^T$, $\bx$ is a fixed design $n \times p$
design matrix, $\beta^0$ is the true underlying $p \times1$ parameter
vector and $\eps$ is the $n \times1$ stochastic error vector with
$\eps_1,\ldots,\eps_n$ i.i.d. having $\EE[\eps_i] = 0$ and $\Var
(\eps_i) =
\sigma^2 < \infty$; throughout the paper, $p$ may be much larger $n$.

We are interested in testing one or many null-hypotheses of the form:
%
\begin{equation}
\label{hypoth} H_{0,G}:\ \beta^0_j = 0 \mbox{
for all } j \in G,
\end{equation}
where $G \subseteq\{1,\ldots,p\}$ is a subset of all the indices of the
covariables. Of substantial interest is the case where $G = \{j\}$
corresponding to a hypothesis for the individual $j$th regression
parameter ($j=1,\ldots,p$). At the other end of the spectrum is the
global null-hypothesis where $G = \{1,\ldots,p\}$, and we allow for any
$G$ between an individual and the global hypothesis.

\subsection{Past work about high-dimensional linear
models}\label{subsec.pastwork}

We review in this section an important stream of research for
high-dimensional linear models. The more familiar reader may skip Section
\ref{subsec.pastwork}.

\subsubsection{The Lasso}

The Lasso \citep{tibs96}
\begin{eqnarray*}
\hat{\beta}_{\mathrm{Lasso}} = \hat{\beta}_{\mathrm
{Lasso}}(\lambda) =
\argmin_{\beta} \bigl( \|\by- \bx\beta\|_2^2/n +
\lambda\|\beta\|_1 \bigr),
\end{eqnarray*}
has become tremendously popular for estimation in high-dimensional linear
models. The three main themes which have been considered in the past are
prediction of the regression surface (and for a new response variable) with
corresponding measure of accuracy
%
\begin{equation}
\label{predict} \bigl\|\bx\bigl(\hat{\beta}_{\mathrm{Lasso}} - \beta^0\bigr)
\bigr\|_2^2/n,
\end{equation}
estimation of the parameter vector whose quality is assessed by
%
\begin{equation}
\label{est-norms} \bigl\|\hat{\beta}_{\mathrm{Lasso}} - \beta^0
\bigr\|_q\ \bigl(q \in\{1,2\}\bigr),
\end{equation}
and variable selection or estimating the support of $\beta^0$, denoted by
the active set $S_0 = \{j;\ \beta^0_j \neq0,\ j=1,\ldots,p\}$ such that
%
\begin{equation}
\label{var-sel} \PP[\hat{S} = S_0]
\end{equation}
is large for a selection (estimation) procedure $\hat{S}$.

\citet{greenrit03} proved the first result closely related to
prediction as
measured in
(\ref{predict}). Without any conditions on the deterministic design matrix
$\bx$, except that the columns are normalized such\vadjust{\goodbreak} that $(n^{-1} \bx^T
\bx)_{jj} \equiv1$, one has with high probability at least $1 - 2
\exp(-
t^2/2)$:
%
\begin{eqnarray}
\label{slow-rate} & &\bigl\|\bx\bigl(\hat{\beta}_{\mathrm{Lasso}}(\lambda) -
\beta^0\bigr)\bigr\|_2^2/n \le3/2 \lambda\bigl\|
\beta^0\bigr\|_1,
\nonumber
\\[-8pt]
\\[-8pt]
\nonumber
& &\lambda= 4 \sigma\sqrt{\frac{t^2 + 2 \log(p)}{n}},
\end{eqnarray}
see \citeauthor{pbvdg11} (\citeyear{pbvdg11}, Cor.~6.1). Thereby, we assume Gaussian errors but
such an
assumption can be relaxed (\citeauthor{pbvdg11}, \citeyear{pbvdg11}, formula (6.5)). From an
asymptotic point of view (where $p$ and $n$ diverge to $\infty$), the
regularization
parameter $\lambda\asymp\sqrt{\log(p)/n}$ leads to consistency for
prediction if the truth is sparse with respect to the $\ell_1$-norm such
that $\|\beta^0\|_1 = o(\lambda^{-1}) = o(\sqrt{n/\log(p)})$. The
convergence rate is then at best $O_P(\lambda) = O_P(\sqrt{\log(p)/n})$
assuming $\|\beta^0\|_1 \asymp1$.

Such a slow rate of convergence can be improved under additional assumptions
on the design matrix $\bx$. The ill-posedness of the design matrix can be
quantified using the\vspace*{1pt} concept of ``modified'' eigenvalues. Consider the
matrix $\hat{\Sigma} = n^{-1} \bx^T \bx$. The smallest eigenvalue of
$\hat{\Sigma}$ is
\[
\lambda_{\mathrm{min}}(\hat{\Sigma}) = \min_{\beta} \beta^T
\hat {\Sigma} \beta.
\]
Of course, $\lambda_{\mathrm{min}}(\hat{\Sigma})$ equals zero if $p >
n$. Instead of taking the minimum on the right-hand side over all $p
\times
1$ vectors $\beta$, we replace it by a \emph{constrained} minimum,
typically over a cone. This leads to the concept of restricted eigenvalues
(\citepp{brt09}; \citeauthor{koltch09a} \citeyear{koltch09b,koltch09a}; \citepp{rasketal10}) or weaker forms such as
the compatibility
constants \citep{vandeGeer:07a} or further slight weakening of the latter
\citep{sunzhang11}. Relations among the different conditions and ``modified''
eigenvalues are discussed in \citet{van2009conditions} and
\citeauthor{pbvdg11} (\citeyear{pbvdg11}, Ch.~6.13). Assuming that the smallest ``modified''
eigenvalue is larger than zero, one can derive an oracle inequality of the
following prototype: with probability at least $1 - 2 \exp(- t^2/2)$ and
using $\lambda$ as in (\ref{slow-rate}):
%
\begin{eqnarray}
\label{oracle-ineq} \bigl\|\bx\bigl(\hat{\beta}_{\mathrm{Lasso}}(\lambda) -
\beta^0\bigr)\bigr\|_2^2/n + \lambda \bigl\|\hat{
\beta}_{\mathrm{Lasso}} - \beta^0\bigr\|_1 \le4
\lambda^2 s_0/\phi_0^2,
\end{eqnarray}
where $\phi_0$ is the compatibility constant (smallest ``modified''
eigenvalue) of the fixed design matrix
$\bx$ (\citeauthor{pbvdg11}, \citeyear{pbvdg11}, Cor.~6.2). Again, this holds by assuming Gaussian
errors but the result can be extended to non-Gaussian distributions. From
(\ref{oracle-ineq}), we have two immediate implications:
from an asymptotic point of view, using $\lambda\asymp\sqrt{\log(p)/n}$
and assuming that $\phi_0$ is bounded away from 0,
%
\begin{eqnarray}
& &\bigl\|\bx\bigl(\hat{\beta}_{\mathrm{Lasso}}(\lambda) - \beta^0\bigr)
\bigr\|_2^2/n = O_P\bigl(s_0 \log(p)/n
\bigr),\label{fast-rate}
\\
& &\bigl\|\hat{\beta}_{\mathrm{Lasso}}(\lambda) - \beta^0\bigr\|_1
= O_P\bigl(s_0 \sqrt{\log(p)/n}\bigr),\label{lassoell1}
\end{eqnarray}
i.e., a fast convergence rate for prediction as in (\ref{fast-rate})
and an
$\ell_1$-norm bound for the estimation error. We note that the oracle
convergence rate, where an oracle would know the active set $S_0$, is
$O_P(s_0/n)$: the $\log(p)$-factor\vadjust{\goodbreak} is the price to pay by not knowing the
active set $S_0$.
An $\ell_2$-norm bound can be derived as well:
$\|\hat{\beta}_{\mathrm{Lasso}}(\lambda) - \beta^0\|_2 = O_P(\sqrt{s_0
\log(p)/n})$ assuming a slightly stronger restricted eigenvalue
condition. Results along these lines have been established by
\citet{buneaetal06}, \citet{geer07} who covers generalized linear
models as well,
\citet{zhang2008sparsity}, \citet{MY08}, \citet{brt09} among others.

The Lasso is doing variable selection: a simple estimator of the active set
$S_0$ is $\hat{S}_{\mathrm{Lasso}}(\lambda) = \{j;\
\hat{\beta}_{\mathrm{Lasso};j}(\lambda) \neq0\}$. In order that
$\hat{S}_{\mathrm{Lasso}}(\lambda)$ has good accuracy for $S_0$, we have
to require
that the non-zero regression coefficients are sufficiently large (since
otherwise, we cannot detect the variables in $S_0$ with high
probability). We make a ``beta-min'' assumption whose asymptotic form reads
as
%
\begin{equation}
\label{beta.min} \min_{j \in S_0} \bigl|\beta_j^0\bigr| \gg
\sqrt{s_0 \log(p)/n}.
\end{equation}
Furthermore, when making a restrictive assumption for the design, called
neighborhood stability, or assuming the equivalent
irrepresentable condition, and choosing a suitable $\lambda\gg
\sqrt{\log(p)/n}$:
\[
\PP\bigl[\hat{S}_{\mathrm{Lasso}}(\lambda) = S_0\bigr] \to1,
\]
see \citet{mebu06}, \citet{zhaoyu06}, and \citet{Wai08} establishes exact
scaling results. The ``beta-min'' assumption in (\ref{beta.min}) as
well as the
irrepresentable condition on the design are restrictive and
non-checkable. Furthermore, these conditions are essentially necessary
(\citepp{mebu06}; \citepp{zhaoyu06}). Thus, under weaker assumptions, we can only derive
a weaker yet useful result about variable screening. Assuming a restricted
eigenvalue condition on the fixed design $\bx$ and the
``beta-min'' condition in (\ref{beta.min}) we still have asymptotically
that for $\lambda\asymp\sqrt{\log(p)/n}$:
%
\begin{equation}
\label{var-screening} \PP\bigl[\hat{S}(\lambda) \supseteq S_0\bigr]
\to1\ (n \to\infty).
\end{equation}
The cardinality of the estimated active set (typically) satisfies
$|\hat{S}(\lambda)| \le\min(n,p)$: thus if $p \gg n$, we achieve a massive
and often useful dimensionality reduction in the original covariates.

We summarize that a slow convergence rate for prediction ``always''
holds. Assuming some ``constrained minimal eigenvalue'' condition on the
fixed design $\bx$, we obtain the fast convergence rate in
(\ref{fast-rate}), and an estimation error bound as in (\ref
{lassoell1}); with
the additional ``beta-min'' assumption, we obtain the practically useful
variable screening property in (\ref{var-screening}). For consistent
variable selection, we necessarily need a (much) stronger condition on
the fixed
design, and such a strong condition is questionable to be true in a
practical problem. Hence variable selection might be a too ambitious goal
with the Lasso. That is why the original translation of Lasso (Least
Absolute Shrinkage and Selection Operator) may be better re-translated
as Least
Absolute Shrinkage and \emph{Screening} Operator. We refer to
\citet{pbvdg11} for an extensive treatment of the properties of the Lasso.

\subsubsection{Other methods}

Of course, the three main inference tasks in a high-dimensional
linear model, as described by~(\ref{predict}), (\ref{est-norms}) and
(\ref{var-sel}), can be pursued with other methods than the
Lasso.

An interesting line of proposals include concave penalty functions
instead of
the $\ell_1$-norm in the Lasso, see for example \citet{fanli01} or
\citet{zhang2010}. The adaptive
Lasso \citep{zou06}, analyzed in the high-dimensional setting by
\citet{huangetal06} and \citet{geer11}, can be interpreted as an
approximation of some concave penalization approach \citep{zouli08}. A
related procedure to the adaptive Lasso is the relaxed Lasso
\citep{Meinshausen:05}. Another method is the Dantzig selector
\citep{cantao07} which has similar statistical properties as the Lasso
\citep{brt09}. Other algorithms include orthogonal matching pursuit (which
is essentially forward variable selection) or
$L_2$Boosting (matching pursuit) which have desirable properties
(\citepp{Tropp04}; \citepp{pb06}).

Quite different from estimation of the high-dimensional parameter vector
are variable screening procedures which aim for an analogous property
as in
(\ref{var-screening}). Prominent examples include the ``Sure Independence
Screening'' (SIS) method
\citep{fanlv07}, and high-dimensional variable screening or selection
properties have been established for forward variable selection
\citep{wang09} and for the PC-algorithm \citep{pbkama09} (``PC''
stands for
the first names of its inventors, Peter Spirtes and Clark Glymour).

\subsection{Assigning uncertainties and $p$-values for high-dimensional
regression}\label{subsec.uncertass}

At the core of statistical inference is the specification of statistical
uncertainties, significance and confidence. For example, instead of having
a variable selection result where the probability in~(\ref{var-sel}) is
large, we would like to have
measures controlling a type I error (false positive selections), including
$p$-values which are adjusted for large-scale multiple testing, or
construction of confidence intervals or regions. In the high-dimensional
setting, answers to these core goals are challenging.

\citet{mebu10} propose Stability Selection, a very generic method
which is
able to control the expected number of false positive selections: that is,
denoting by $V = |\hat{S} \cap S_0^c|$, Stability Selection yields a
finite-sample upper bound of $\EE[V]$ (not only for linear models but
also for
many other inference problems). To achieve this, a very restrictive
(but presumably non-necessary) exchangeability condition is made which, in
a linear model, is
implied by a restrictive assumption for the design matrix. On the positive
side, there is no requirement of a ``beta-min'' condition as in
(\ref{beta.min}) and the method seems to provide reliable control of
$\EE[V]$.

\citet{WR08} propose a procedure for variable selection based on sample
splitting. Using their idea and extending it to multiple sample splitting,
\citet{memepb09}
develop a much more stable method for construction of
$p$-values for hypotheses $H_{0,j}:\ \beta^0_j =0\ (j=1,\ldots,p)$ and for
adjusting them in a non-naive way for multiple testing over $p$ (dependent)
tests. The main drawback of this procedure is its required ``beta-min''
assumption in (\ref{beta.min}). And this is very undesirable since for
statistical hypothesis testing, the test should control type I error
regardless of the size of the coefficients, while the power of the test
should be large if the absolute value of the coefficient would be large:
thus, we should avoid assuming (\ref{beta.min}).

Up to now, for the high-dimensional linear model case with $p
\gg n$, it seems that only \citet{zhangzhang11} managed to construct a
procedure which leads to statistical tests for $H_{0,j}$ without
assuming a
``beta-min'' condition.

\subsection{A loose description of our new results}

Our starting point is Ridge regression for estimating the high-dimensional
regression parameter. We then develop a bias correction,
addressing the issue that Ridge regression is estimating the regression
coefficient vector projected to the row space of the design matrix: the
corrected estimator is denoted by $\hat{\beta}_{\mathrm{corr}}$.

Theorem
\ref{th1} describes that under the null-hypothesis, the distribution
of a
suitably normalized $a_{n,p} |\hat{\beta}_{\mathrm{corr}}|$ can be
asymptotically and stochastically (componentwise) upper-bounded:
%
\begin{eqnarray}
\label{formula-descr} & &a_{n,p} |\hat{\beta}_{\mathrm{corr}}| \stackrel{
\mathrm {as}} {\preceq} \bigl(|Z_j| + \Delta_j\bigr)_{j=1}^p,
\nonumber
\\[-8pt]
\\[-8pt]
\nonumber
& &(Z_1,\ldots,Z_p) \sim\mathcal{ N}_p
\bigl(0,\sigma^2 n^{-1} \Omega\bigr),
\end{eqnarray}
for some \emph{known} positive definite matrix $\Omega$ and some
\emph{known} constants $\Delta_j$. This is the key to derive $p$-values based
on this stochastic upper bound. It can be used for construction of $p$-values
for individual hypotheses
$H_{0,j}$ as well as for more global hypotheses $H_{0,G}$ for
\emph{any} subset $G \subseteq\{1,\ldots,p\}$, including cases where $G$
is (very) large. Furthermore, Theorem~\ref{th2} justifies a simple approach
for controlling the familywise error rate when considering multiple testing
of regression hypotheses. Our multiple testing adjustment method itself is
closely related to the Westfall--Young permutation procedure
\citep{westyoung93} and hence, it offers high power, especially in presence
of dependence among the many test-statistics \citep{memabu11}.

\subsubsection{Relation to other work}\label{subsec.otherwork}

Our new method as well as the approach in \citet{zhangzhang11} provide
$p$-values (and the latter also confidence intervals) without assuming a
``beta-min'' condition. Both of them build on using linear estimators
and a
correction using a non-linear initial estimator such as the Lasso. Using
e.g., the Lasso directly leads to the problem of characterizing the
distribution of the estimator (in a tractable form): this seems very
difficult in
high-dimensional settings while it has been worked out for low-dimensional
problems \citep{knfu00}.
The work by \citet{zhangzhang11} is the only one which studies (sufficiently
closely) related questions and goals as in this paper.

The approach by \citet{zhangzhang11} is based on the idea of
projecting the
high-dimensional parameter vector to low-dimensional components, as
occurring naturally in the hypotheses $H_{0,j}$ about single
components, and then proceeding with a linear estimator. This idea is pursued
with the ``efficient score function''
approach from semiparametric statistics \citep{bicketal98}. The
difficulty in the high-dimensional setting is the construction of the
score vector $z_j$ from which one can derive a confidence interval for
$\beta^0_j$: \citet{zhangzhang11} propose it as the residual vector from
the Lasso when regressing $\bx^{(j)}$ against all other variables
$\bx^{(\setminus j)}$
(where $\bx^{(J)}$ denotes the design sub-matrix whose columns correspond
to the index set $J \subseteq\{1,\ldots,p\}$). They then prove
the asymptotic validity of confidence intervals for finite, sparse linear
combinations of $\beta^0$. The difference to our work is primarily a rather
different construction of
the projection where we make use of Ridge estimation with a very simple
choice of regularization. A drawback of our method is that, typically,
it is
not theoretically rate-optimal in terms of power.

\section{Model, estimation and $p$-values}
Consider one or many null-hypotheses as in
(\ref{hypoth}). We are interested in constructing $p$-values for hypotheses
$H_{0,G}$ without imposing a ``beta-min'' condition as in (\ref{beta.min}):
the statistical test itself will distinguish whether a regression
coefficient is small or not.

\subsection{Identifiability}\label{subsec.identif}
We consider model (\ref{mod.lin}) with fixed design. Without making
additional assumptions on the design matrix $\bx$, there is a problem of
identifiability. Clearly, if $p > n$ and hence $\operatorname{rank}(\bx) \le
n <
p$, there are different parameter vectors $\theta$ such that
$\bx\beta^0 = \bx\theta$. Thus, we cannot identify $\beta^0$ from the
distribution of $Y_1,\ldots,Y_n$ (and fixed design $\bx$).

\citet{shadeng11} give a characterization of identifiability in a
high-dimensional linear model (\ref{mod.lin}) with fixed design. Following
their approach, it is useful to consider the singular value decomposition
\begin{eqnarray*}
& &\bx= RSV^T,
\\
& &R \mbox{ $n \times n$ matrix with}\ R^T R = I_n,
\\
& &S \mbox{ $n \times n$ diagonal matrix with singular values}\
s_1,\ldots ,s_n,
\\
& &V \mbox{ $p \times n$ matrix with}\ V^T V = I_n.
\end{eqnarray*}
Denote by $\mathcal{ R}(\bx) \subset
\R^p$ the linear space generated by the $n$ rows of $\bx$. The projection
of $\R^p$ onto $\mathcal{ R}(\bx)$ is then
\[
P_{\bx} = \bx^T \bigl(\bx\bx^T
\bigr)^{-} \bx= V V^T,
\]
where $A^{-}$ denotes the pseudo-inverse of a squared matrix $A$.

A natural choice of a parameter $\theta^0$ such that $\bx\beta^0 =
\bx\theta^0$
is the projection of $\beta^0$ onto $\mathcal{ R}(\bx)$. Thus,
%
\begin{equation}
\label{theta} \theta^0 = P_{\bx} \beta^0 = V
V^T \beta^0.
\end{equation}
Then, of course, $\beta^0 \in\mathcal{ R}(\bx)$ if and only if
$\beta^0 =
\theta^0$.

\subsection{Ridge regression}\label{subsec.Ridgeplarge}

Consider Ridge regression
%
\begin{eqnarray}
\label{Ridgetheta} \hat{\beta} = \operatorname{argmin}_{\beta} \|\by- \bx\beta
\|_2^2/n + \lambda \|\beta\|_2^2 =
\bigl(n^{-1} \bx^T \bx+ \lambda I_p
\bigr)^{-1} n^{-1} \bx^T \by,
\end{eqnarray}
where $\lambda= \lambda_n$ is a regularization parameter. By construction
of the estimator, $\hat{\beta} \in\mathcal{ R}(\bx)$; and indeed, as
discussed below, $\hat{\beta}$ is a reasonable estimator for $\theta^0 = P_\mathbf{X}
\beta^0$. We denote by
\[
\hat{\Sigma} = n^{-1} \bx^T \bx.
\]
The covariance matrix of the Ridge estimator, multiplied by $n$, is then
%
\begin{eqnarray}
\label{omegasvd} \Omega= \Omega(\lambda) &=& (\hat{\Sigma} +
\lambda_n I)^{-1} \hat{\Sigma} (\hat{\Sigma} +
\lambda_n I)^{-1}
\nonumber
\\[-8pt]
\\[-8pt]
\nonumber
& =& V \operatorname{diag}\biggl(\frac{s_1^2}{(s_1^2 + \lambda)^2},\ldots ,\frac{s_n^2}{(s_n^2 + \lambda)^2}\biggr)
V^T,
\end{eqnarray}
a quantity which will appear at many places again. We assume that
%
\begin{equation}
\label{minvar} \Omega_{\mathrm{min}}(\lambda) := \min_{j \in\{1,\ldots,p\}}
\Omega_{jj}(\lambda) > 0.
\end{equation}
We do not require that $\Omega_{\mathrm{min}}(\lambda)$ is bounded away
from zero as a function of $n$ and $p$. Thus, the assumption in
(\ref{minvar}) is very mild: a rather
peculiar design would be needed to violate the condition, see also the
equivalent formulation in formula (\ref{minvar2}) below. Furthermore,
(\ref{minvar}) is easily checkable.

We denote by $\lambda_{\mathrm{min} \neq0}(A)$ the smallest non-zero
eigenvalue of a symmetric matrix $A$. We then have the following result.
%
\begin{prop}\label{prop1}
Consider the Ridge regression estimator $\hat{\beta}$ in
(\ref{Ridgetheta}) with regularization parameter
$\lambda> 0$. Assume condition (\ref{minvar}), see also
(\ref{minvar2}).
Then,
\begin{eqnarray*}
& &\max_{j \in\{1,\ldots,p\}}\bigl|\EE[\hat{\beta}_j ] - \theta^0_j\bigr|
\le \lambda\bigl\|\theta^0\bigr\|_2 \lambda_{\mathrm{min} \neq0}(\hat{
\Sigma })^{-1},
\\
& &\min_{j \in\{1,\ldots,p\}} \Var(\hat{\beta}_j) \ge n^{-1}
\sigma^2 \Omega_{\mathrm{min}}(\lambda).
\end{eqnarray*}
\end{prop}

A proof is given in Section~\ref{sec.proofs}, relying in
large parts on
\citet{shadeng11}. We now discuss under which circumstances the estimation
bias is smaller than
the standard error. Qualitatively, this happens if $\lambda>0$ is chosen
sufficiently small. For a more quantitative discussion, we study the
behavior of $\Omega_{\mathrm{min}}(\lambda)$ as a function of
$\lambda$ and
we obtain an equivalent formulation of~(\ref{minvar}).
%
\begin{lemm}\label{lemm1}
We have the following:
\begin{enumerate}
\item
\[
\Omega_{\mathrm{min}}(\lambda) = \min_j \sum
_{r=1}^n \frac{s_r^2}{(s_r^2
+ \lambda)^2} V_{jr}^2.
\]
From this we get:
%
\begin{equation}
\label{minvar2} \mbox{(\ref{minvar}) holds if and only if }\min_{1 \le j \le p}
\max_{1 \le r \le n, s_r \neq0} V_{jr}^2 > 0.
\end{equation}
\item
Assuming (\ref{minvar}),
\[
\Omega_{\mathrm{min}}\bigl(0^+\bigr) := \lim_{\lambda\searrow0^+}
\Omega_{\mathrm{min}}(\lambda) = \min_j \sum
_{r= 1;s_r \neq0}^n \frac{1}{s_r^2}V_{jr}^2
> 0.
\]
\item
%
\begin{equation}
\label{omegamin} \mbox{if (\ref{minvar}) holds: } 0 < L_C \le
\liminf_{\lambda\in(0,C]} \Omega_{\mathrm{min}}(\lambda) \le M_C <
\infty,
\end{equation}
for any $0 < C < \infty$, and where $0 < L_C < M_C < \infty$ are constants
which depend on $C$ and on the design matrix $\bx$ (and hence on $n$ and
$p$).
\end{enumerate}
\end{lemm}
The proof is straightforward using the expression (\ref{omegasvd}). The
statement 3. says that for a given data-set, the variances of the
$\hat{\beta}_j$'s remain in a reasonable range even if we choose
$\lambda>
0$ arbitrarily small; the statement doesn't imply anything for the behavior
as $n$ and $p$ are getting large (as the data and design matrix change).
From Proposition~\ref{prop1}, we immediately obtain the following result.
%
\begin{corr}
Consider the Ridge regression estimator $\hat{\beta}$ in
(\ref{Ridgetheta}) with regularization parameter
$\lambda> 0$ satisfying
%
\begin{equation}
\label{lambdachoice} \lambda\Omega_{\mathrm{min}}(\lambda)^{-1/2} \le
n^{-1/2} \sigma\bigl\|\theta^0\bigr\|_2^{-1}
\lambda_{\mathrm{min} \neq
0}(\hat{\Sigma}).
\end{equation}
In addition, assume condition (\ref{minvar}), see also
(\ref{minvar2}).
Then
\[
\max_{j \in\{1,\ldots,p\}}\bigl(\EE[\hat{\beta}_j ] -
\theta^0_j\bigr)^2 \le\min_{j
\in\{1,\ldots,p\}}
\Var(\hat{\beta}_j).
\]
Due to the third statement in Lemma~\ref{lemm1} regarding the behavior of
$\Omega_{\mathrm{min}}(\lambda)$, (\ref{lambdachoice}) can be
fulfilled for
a sufficiently small value of $\lambda$ (a more precise
characterization of
the maximal $\lambda$ which fulfills~(\ref{lambdachoice}) would require
knowledge of $\|\theta^0\|_2$).
\end{corr}

\subsection{The projection bias and corrected Ridge
regression}\label{subsec.projbias}

As discussed in Section~\ref{subsec.identif}, Ridge regression is
estimating the parameter $\theta^0 = P_{\bx} \beta^0$ given in~(\ref{theta}).
Thus, in general, besides the estimation bias governed by
the choice of $\lambda$, there is an additional projection bias $B_j =
\theta^0_j - \beta^0_j\ (j=1,\ldots,p)$.
Clearly,
\begin{eqnarray*}
B_j = \bigl(P_{\bx} \beta^0
\bigr)_j - \beta^0_j = (P_{\bx})_{jj}
\beta^0_j - \beta^0_j + \sum
_{k
\neq j} (P_{\bx})_{jk}
\beta^0_k.
\end{eqnarray*}
In terms of constructing $p$-values, controlling type I error for testing
$H_{0,j}$ or $H_{0,G}$ with $j \in G$, the projection bias has only a
disturbing effect if $\beta^0_j = 0$ and $\theta^0_j \neq0$, and we
only have
to consider the bias under the null-hypothesis:
%
\begin{equation}
\label{hoj} B_{H_0;j} = \sum_{k \neq j}
(P_{\bx})_{jk} \beta^0_k.
\end{equation}
The bias $B_{H_0;j}$ is also the relevant quantity for the case under the
non null-hypothesis, see the brief comment after Proposition
\ref{prop-repr}.
We can estimate $B_{H_0;j}$ by
\[
\hat{B}_{H_0;j} = \sum_{k \neq j}
(P_{\bx})_{jk} \hat{\beta }_{\mathrm{init};k},
\]
where $\hat{\beta}_{\mathrm{init}}$ is an initial estimator such as the
Lasso which guarantees a certain estimation accuracy, see assumption \textup{(A)}
below.
This motivates the following bias-corrected Ridge estimator for testing
$H_{0,j}$, or $H_{0,G}$ with $j \in G$:
%
\begin{equation}
\label{Ridgecorr} \hat{\beta}_{\mathrm{corr};j} = \hat{\beta}_j -
\hat{B}_{H_0;j} = \hat{\beta}_j - \sum
_{k \neq j} (P_{\bx})_{jk} \hat{\beta
}_{\mathrm{init};k}.
\end{equation}
We then have the following representation.
%
\begin{prop}\label{prop-repr}
Assume model (\ref{mod.lin}) with Gaussian errors. Consider the corrected
Ridge regression estimator $\hat{\beta}_{\mathrm{corr}}$ in
(\ref{Ridgecorr}) with regularization parameter $\lambda> 0$,
and assume (\ref{minvar}). Then,
\begin{eqnarray*}
& &\hat{\beta}_{\mathrm{corr};j} = Z_j + \gamma_j\ (j=1,
\ldots,p)
\\
& &Z_1,\ldots,Z_p \sim\mathcal{ N}_p
\bigl(0,n^{-1} \sigma^2 \Omega\bigr),\ \Omega= \Omega(
\lambda),
\\
& &\gamma_j = (P_{\bx})_{jj}
\beta^0_j - \sum_{k \neq j}
(P_{\bx
})_{jk}\bigl(\hat{\beta}_{\mathrm{init};k} -
\beta^0_k\bigr) + b_j(\lambda ),
\\
& &b_j(\lambda) = \EE\bigl[\hat{\beta}_j(\lambda)
\bigr] - \theta^0_j.
\end{eqnarray*}
\end{prop}

A proof is given in Section~\ref{sec.proofs}. We infer from
Proposition~\ref{prop-repr} a representation which could be used not only
for testing but also for constructing confidence intervals:
\begin{eqnarray*}
\frac{\hat{\beta}_{\mathrm{corr};j}}{(P_{\bx})_{jj}} - \beta_j^0 = \frac{Z_j}{(P_{\bx})_{jj}} -
\sum_{k \neq j} \frac{(P_{\bx})_{jk}}{(P_{\bx})_{jj}}\bigl(\hat{
\beta}_{\mathrm
{init};k} - \beta^0_k\bigr) +
\frac{b_j(\lambda)}{(P_{\bx})_{jj}}.
\end{eqnarray*}
The normalizing
factors for the
variables $Z_j$ bringing them to the $\mathcal{ N}(0,1)$-scale are
\[
a_{n,p;j}(\sigma) = n^{1/2} \sigma^{-1}
\Omega_{jj}^{-1/2}\ (j=1,\ldots,p)
\]
which are also depending on $\lambda$ through $\Omega=
\Omega(\lambda)$. We refer to Section~\ref{subsec.anrate} where the
unusually fast divergence of $a_{n,p;j}(\sigma)$ is discussed. The
test-statistics we consider are simple functions of $a_{n,p;j}(\sigma)
\hat{\beta}_{\mathrm{corr};j}$.

\subsection{Stochastic bound for the distribution of the corrected Ridge
estimator: Asymptotics}

We provide here an asymptotic stochastic bound for the distribution of
$a_{n,p;j}(\sigma) \hat{\beta}_{\mathrm{corr};j}$ under the
null-hypothesis. The asymptotic formulation is compact and the basis for
the construction of $p$-values in
Section~\ref{sec.pvalues}, but we give more detailed finite-sample results
in Section~\ref{sec.finites}.

We consider a triangular array of observations from a linear model as in
(\ref{mod.lin}):
%
\begin{equation}
\label{mod.lin2} \by_n = \bx_n \beta^0_n
+ \eps_n,\ n=1,2,\ldots,
\end{equation}
where all the quantities and also the dimension $p = p_n$ are allowed to
change with $n$. We make the following assumption.
\begin{description}
\item[(A)] There are constants $\Delta_j = \Delta_{j,n} > 0$ such that
\begin{eqnarray*}
\PP\Biggl[\bigcap_{j=1}^{p_n} \biggl\{\biggl|a_{n,p;j}(\sigma)
\sum_{k \neq j} (P_{\bx
})_{jk}\bigl(
\hat{\beta}_{\mathrm{init};k} - \beta^0_k\bigr)\biggr| \le
\Delta_{j,n}\biggr\}\Biggr] \to 1\ (n \to\infty).
\end{eqnarray*}
\end{description}
We will discuss in Section~\ref{subsec.Delta} constructions for such bounds
$\Delta_j$ (which are typically not negligible).
Our next result is the key to obtain a $p$-value for testing the
null-hypothesis $H_{0,j}$ or $H_{0,G}$, saying that asymptotically,
\[
a_{n,p;j}(\sigma) |\hat{\beta}_{\mathrm{corr};j}| \stackrel{\mathrm{as.}} {
\preceq} |W| + \Delta_j,
\]
where $W \sim\mathcal{ N}(0,1)$, and similarly for the multi-dimensional
version with $\hat{\beta}_{\mathrm{corr};G}$ (where $\preceq$ denotes
``stochastically smaller or equal to'').
%
\begin{theo}\label{th1}
Assume model (\ref{mod.lin2}) with fixed design and Gaussian
errors. Consider the corrected
Ridge regression estimator $\hat{\beta}_{\mathrm{corr}}$ in (\ref
{Ridgecorr}) with regularization parameter $\lambda_n > 0$ such that
\begin{eqnarray*}
\lambda_n \Omega_{\mathrm{min}}(\lambda_n)^{-1/2}
= o\bigl(\min\bigl(n^{-1/2} \bigl\|\theta^0\bigr\|_2^{-1}
\lambda_{\mathrm{min} \neq0}(\hat{\Sigma})\bigr)\bigr)\ (n \to\infty),
\end{eqnarray*}
and assume condition \textup{(A)} and (\ref{minvar}) (while
for the latter, the quantity does not need to be bounded away from zero).
Then, for $j \in\{1,\ldots,p_n\}$ and if $H_{0,j}$ holds: for all $u
\in
\R^+$,
\begin{eqnarray*}
\limsup_{n \to\infty} \bigl(\PP\bigl[a_{n,p;j}(\sigma) |\hat{\beta
}_{\mathrm{corr};j}| > u\bigr] - \PP\bigl[|W| + \Delta_j > u\bigr] \bigr) \le0,
\end{eqnarray*}
where $W \sim\mathcal{ N}(0,1)$.
Similarly, for any sequence of subsets $\{G_n\}_n,\ G_n \subseteq
\{1,\ldots,p_n\}$ and if $H_{0,G_n}$ holds: for all $u \in\R^+$,
\begin{eqnarray*}
\limsup_{n \to\infty} \Bigl(\PP\Bigl[\max_{j \in G_n} a_{n,p;j}(
\sigma) |\hat{\beta}_{\mathrm{corr};j}| > u\Bigr] - \PP\Bigl[\max_{j
\in G_n}
\bigl(a_{n,p;j}(\sigma) |Z_j| + \Delta_j\bigr) >
u\Bigr] \Bigr) \le0,
\end{eqnarray*}
where $Z_1,\ldots,Z_P$ are as in Proposition~\ref{prop-repr}.
\end{theo}

{\spaceskip=0.19em plus 0.05em minus 0.02em  A proof is given in Section~\ref{sec.proofs}. As written
above already, due to the third statement in Lemma~\ref{lemm1},} the
condition for $\lambda_n$ is reasonable. We note that the
distribution of $\max_{j \in G_n} (a_{n,p;j}(\sigma) |Z_j| + \Delta_j)$
does not depend on $\sigma$ and can
be easily computed via simulation.

\subsubsection{\texorpdfstring{Bounds $\Delta_j$ in assumption \textup{(A)}}
{Bounds Delta j in assumption (A)}}\label{subsec.Delta}

We discuss an approach for constructing the bounds $\Delta_j$. As
mentioned above, they should not involve any unknown quantities so that we
can use them for constructing $p$-values from the distribution of
$|W| + \Delta_j$ or $\max_{j \in G_n} (a_{n,p;j}(\sigma)
|Z_j| + \Delta_j)$, respectively.

We rely on the (crude) bound
%
\begin{eqnarray}
\label{crudebound} \biggl|a_{n,p;j}(\sigma) \sum_{k \neq j}
(P_{\bx})_{jk}\bigl(\hat{\beta }_{\mathrm{init};k} -
\beta^0_k\bigr)\biggr| \le a_{n,p;j}(\sigma)
\max_{k \neq j}\bigl|(P_{\bx})_{jk}\bigr| \bigl\|\hat{
\beta}_{\mathrm{init}} - \beta^0\bigr\|_1.
\end{eqnarray}
To proceed further, we consider the Lasso as initial estimator. Due to
(\ref{oracle-ineq}) we obtain
%
\begin{eqnarray}
\label{crudebound2}\biggl |a_{n,p;j}(\sigma) \sum_{k \neq j}
(P_{\bx})_{jk}\bigl(\hat{\beta }_{\mathrm{init};k} -
\beta^0_k\bigr)\biggr| \le\max_{k \neq j}
\bigl|a_{n,p;j}(\sigma) (P_{\bx})_{jk}\bigr| 4
\lambda_{\mathrm{Lasso}} s_0 \phi_0^{-2},
\end{eqnarray}
where the last inequality holds on a set with probability at least $1
- 2\exp(-t^2/2)$ when choosing $\lambda_{\mathrm{Lasso}}$ as in
(\ref{slow-rate}). The assumptions we require are summarized next.
%
\begin{lemm}\label{lemm.bound}
Consider the linear model (\ref{mod.lin2}) with fixed design, having
normalized columns $\hat{\Sigma}_{jj} \equiv1$, which satisfies
the compatibility condition with constant $\phi_0^2 =
\phi_{0,n}^2$. Consider the Lasso as initial estimator
$\hat{\beta}_{\mathrm{init}}$ with regularization parameter
$\lambda_{\mathrm{Lasso}} = 4 \sigma\sqrt{C\log(p_n)/n}$ for some
$2 < C <
\infty$.
Assume that the sparsity $s_0 = s_{0,n} = \break o((n/\log(p_n))^{\xi})\ (n
\to
\infty)$ for some $0 < \xi< 1/2$, and that $\liminf_{n \to\infty}
\phi_{0,n}^2 > 0$. Then,
%
\begin{equation}
\label{bound1} \Delta_j :\equiv\max_{k \neq j}
\bigl|a_{n,p;j}(\sigma) (P_{\bx})_{jk}\bigr|\bigl(\log(p)/n
\bigr)^{1/2 - \xi}
\end{equation}
satisfies assumption \textup{(A)}.
\end{lemm}

A proof follows from (\ref{crudebound2}). We summarize the results as
follows.
%
\begin{corr}
Assume the conditions of Theorem~\ref{th1} without condition \textup{(A)} and
the conditions of Lemma~\ref{lemm.bound}. Then, when using the Lasso
as initial estimator, the statements in Theorem~\ref{th1} hold.
\end{corr}

The construction of the bound in (\ref{bound1}) requires the compatibility
condition on the design and an upper bound for the sparsity $s_0$.
While the
former is an identifiability condition, and some form of identifiability
assumption is certainly necessary, the latter condition about knowing the
magnitude of the sparsity is not very elegant. When assuming bounded sparsity
$s_{0,n} \le M < \infty$ for all $n$, we can choose $\xi= 0$ with
an additional constant $M$ on the right-hand side of (\ref{bound1}).
In our
practical examples in Section~\ref{sec.numeric}, we use $\xi= 0.05$.

\subsection{$P$-values}\label{sec.pvalues}
Our construction of $p$-values is based on the asymptotic distributions in
Theorem~\ref{th1}. For an individual hypothesis $H_{0,j}$, we define the
$p$-value for the two-sided alternative as
%
\begin{equation}
\label{pvalue1} P_j = 2 \bigl(1 - \Phi \bigl(\bigl(a_{n,p;j}(
\sigma) |\hat{\beta}_{\mathrm{corr};j}| - \Delta_j\bigr)_+ \bigr)
\bigr).
\end{equation}
Of course, we could also consider one-sided alternatives with the obvious
modification for $P_j$. For a more general hypothesis $H_{0,G}$ with
$|G| >
1$, we use the maximum as test statistics (but other statistics such as
weighted sums could be chosen as well) and denote by
\begin{eqnarray*}
& &\hat{\gamma}_{G} = \max_{j \in G} a_{n,p;j}(\sigma) |
\hat{\beta}_{\mathrm{corr};j}|,
\\
& &J_G(c) = \PP\Bigl[\max_{j \in G} \bigl(a_{n,p;j}(
\sigma) |Z_j| + \Delta_j\bigr) \le c\Bigr],
\end{eqnarray*}
where the latter is independent of $\sigma$ and can be easily computed via
simulation ($Z_1,\ldots,Z_p$ are as in Proposition~\ref{prop-repr}).
Then, the $p$-value for $H_{0,G}$, against the alternative being the
complement $H_{0,G}^c$, is defined as
%
\begin{equation}
\label{pvalue2} P_{G} = 1 - J_G(\hat{
\gamma}_{G}).
\end{equation}
We note that when $\Delta_j \equiv\Delta$ is the same for all $j$,
we can
rewrite $P_G = 1 - \PP[\max_{j \in G} a_{n,p;j}(\sigma)\times |Z_j| \le
(\hat{\gamma}_{G} -
\Delta)_+]$ which is a direct analogue of (\ref{pvalue1}).

Error control follows immediately by the construction of the $p$-values.
%
\begin{corr}\label{corr.pvalue}
Assume the conditions in Theorem~\ref{th1}. Then, for any $0 < \alpha
< 1$,
\begin{eqnarray*}
& &\limsup_{n \to\infty} \PP[P_j \le\alpha] - \alpha\le0\ \mbox
{if $H_{0,j}$ holds},
\\
& &\limsup_{n \to\infty} \PP[P_G \le\alpha] - \alpha\le0\ \mbox
{if $H_{0,G}$ holds}.
\end{eqnarray*}
Furthermore, for any sequence $\alpha_n \to0\ (n \to\infty)$ which
converges sufficiently slowly, the statements also hold when replacing
$\alpha$ by $\alpha_n$.
\end{corr}
A discussion about detection power of the method is given in Section
\ref{subsec.detection}. Further remarks about these $p$-values are given in
Section~\ref{subsec.outlookbound}.

\subsubsection{\texorpdfstring{Estimation of $\sigma$}{Estimation of sigma}}

In practice, for the $p$-values in (\ref{pvalue1}) and (\ref{pvalue2}), we
use the normalizing factor $a_{n,p;j}(\hat{\sigma})$ with an estimate
$\hat{\sigma}$. These $p$-values are asymptotically controlling the
type I
error if $\PP[\hat{\sigma} \ge\sigma] \to1\ (n \to\infty)$.
This follows
immediately from the construction.

We propose to use the estimator $\hat{\sigma}$ from the Scaled Lasso
method \citep{sunzhang11}. Assuming $s_{0} \log(p)/n = o(1)\ (n \to
\infty)$ and the compatibility condition for the design, \citet{sunzhang11}
prove that $|\hat{\sigma}/\sigma- 1| = o_P(1)\ (n \to\infty)$.

\section{Multiple testing}\label{sec.multtest}

We aim to strongly control the familywise error rate $\PP[V>0]$ where $V$
is the
number of false positive selections. For
simplicity, we consider first individual
hypotheses $H_{0,j}\ (j \in\{1,\ldots,p\})$. The generalization to
multiple testing of general hypotheses $H_{0,G}$ with
$|G| > 1$ is discussed in Section~\ref{subsec.multtestgen}.

Based on the individual $p$-values $P_j$, we want to construct corrected
$p$-values $P_{\mathrm{corr};j}$ corresponding to the following decision
rule:
\begin{eqnarray*}
\mbox{reject $H_{0,j}$ if $P_{\mathrm{corr};j} \le\alpha$}\ (0 < \alpha<
1).
\end{eqnarray*}
We denote the associated estimated set of rejected hypotheses (the set of
significant variables) by $\hat{S}_{\alpha} = \{j;\ P_{\mathrm
{corr};j} \le
\alpha\}$. Furthermore,\vspace*{1pt} recall that $S_0 = \{j;\ \beta^0_j \neq0\}$
is the
set of true active variables. The number of false positives using the
nominal significance level $\alpha$ is the denoted by
\[
V_{\alpha} = \hat{S}_{\alpha} \cap S_0^c.
\]
The goal is to construct $P_{\mathrm{corr};j}$ such that $\PP
[V_{\alpha} >
0] \le\alpha$, or that the latter holds at least in an asymptotic sense.
The method we describe here is
closely related to the Westfall--Young procedure \citep{westyoung93}.

Consider the variables $Z_1,\ldots,Z_p \sim\mathcal{ N}_p(0,\sigma^2 n^{-1}
\Omega)$ appearing in Proposition~\ref{prop-repr} or Theorem~\ref
{th1}. Consider the
following distribution function:
\[
F_Z(c) = \PP\Bigl[\min_{1 \le j \le p} 2\bigl(1 - \Phi
\bigl(a_{n,p;j}(\sigma) |Z_j|\bigr)\bigr) \le c\Bigr]
\]
and define
%
\begin{equation}
\label{pcorr} P_{\mathrm{corr};j} = F_Z(P_j + \zeta),
\end{equation}
where $\zeta>0$ is an arbitrarily small number, e.g. $\zeta= 0.01$
for using
the method in practice. Regarding the choice of $\zeta= 0$ (which we use
in all empirical examples in Section~\ref{sec.numeric}), see the Remark
appearing after Theorem~\ref{th2} below.
The distribution function $F_Z(\cdot)$ is independent of $\sigma$ and can
be easily computed via simulation of
the dependent, mean zero jointly Gaussian variables $Z_1,\ldots,Z_p$. It
is computationally (much) faster than simulation of the so-called
minP-statistics \citep{westyoung93} which would require fitting
$\hat{\beta}_{\mathrm{corr}}$ many times.

\subsection{Asymptotic justification of the multiple testing procedure}

We first derive familywise error control in an asymptotic
sense. For a finite sample result, see Section~\ref{sec.finites}.
We consider the
framework as in (\ref{mod.lin2}).
%
\begin{theo}\label{th2}
Assume the conditions in Theorem~\ref{th1}. For the $p$-value in
(\ref{pvalue1}) and using the correction in (\ref{pcorr}) with $\zeta
>0$ we
have: for $0 < \alpha< 1$,
\[
\limsup_{n \to\infty} \PP[V_{\alpha}>0] \le\alpha.
\]
Furthermore, for any sequence $\alpha_n \to0\ (n \to\infty)$ which
converges sufficiently slowly, it holds that $\limsup_{n \to\infty}
\PP[V_{\alpha_n}>0] - \alpha_n \le0$.
\end{theo}

A proof is given in Section~\ref{sec.proofs}.

\begin{rem*}[(Multiple testing correction in \textup{(\protect\ref{pcorr})} with $\bolds{\zeta=0}$)]
We could modify the correction in~(\ref{pcorr}) using $\zeta=0$: the statement in Theorem~\ref{th2}
can then
be derived
when making the additional assumption that
%
\begin{equation}
\label{derivGZ} \sup_{n \in\Nat} \sup_{u} \bigl|F'_{n,Z}(u)\bigr|
< \infty,
\end{equation}
where $F_{n,Z}(\cdot) = F_Z(\cdot)$ is the distribution function appearing
in (\ref{pcorr}) which depends in the asymptotic framework on $n$ and
(mainly on) $p = p_n$. Verifying (\ref{derivGZ}) may not be easy for
general matrices $\Omega= \Omega_{n,p_n}$. However, for the special case
where $Z_1,\ldots,Z_p$ are independent,
\begin{eqnarray*}
F'_Z(u) = p \varphi(u) \bigl(1 - \Phi(u)
\bigr)^{p-1}
\end{eqnarray*}
which is nicely bounded as a function of $u$, over all values of $p$.
\end{rem*}

\subsection{Multiple testing of general hypotheses}\label{subsec.multtestgen}

The methodology for testing many general hypotheses $H_{0,G_j}$ with $|G_j|
\ge1$, $j=1,\ldots,m$ is the same as before. Denote by $S_{0,G} = \{
j;\
H_{0,G_j}\ \mbox{does not hold}\}$ and by $S_{0,G}^c = \{j;\
H_{0,G_j}\
\mbox{holds}\}$; note that these sets are determined by the true parameter
vector~$\beta^0$. Since the $p$-value in (\ref{pvalue2}) is of the form
$P_{G_j} = 1 -
J_{G_j}(\hat{\gamma}_{G_j})$, we consider
\begin{eqnarray*}
F_{G,Z} = \PP\Bigl[\min_{j =1,\ldots,m} \bigl(1 - J_{G_j}(
\gamma_{G_j,Z})\bigr) \le c\Bigr],\ \gamma_{G,Z} =
\max_{j \in G} \bigl(a_{n,p;j}(\sigma) |Z_j|\bigr)
\end{eqnarray*}
which can be easily computed via simulation (and it is independent of
$\sigma$). We then define the corrected $p$-value
as
\[
P_{\mathrm{corr};G_j} = F_{G,Z} (P_{G_j} + \zeta),
\]
where $\zeta> 0$ is a small value such as $\zeta= 0.01$; see also the
definition in (\ref{pcorr}) and the corresponding discussion for the case
where $\zeta=0$ (which now applies to the distribution function
$F_{G,Z}$ instead of $F_Z$).
We denote by
$\hat{S}_{G,\alpha} = \{j;\ P_{\mathrm{corr};G_j} \le
\alpha\}$ and \mbox{$V_{G,\alpha} = \hat{S}_{G,\alpha} \cap S_{0,G}^c$}.

If $J_{G_j}(\cdot)$ has a bounded first derivative, for all $j$, we
can obtain
the same result, under the same conditions, as in Theorem~\ref{th2} (and
without making a condition on the cardinalities of $G_j$). If
$J_{G_j}(\cdot)$
has not a bounded first derivative, we can get around this problem by
modifying the $p$-value $P_{G_j}$ in (\ref{pvalue2}) to $\tilde
{P}_{G_j} = 1 -
J_{G_j}(\hat{\gamma}_{G_j} - \nu)$ for
any (small) $\nu>0$ and proceeding with~$\tilde{P}_{G_j}$.

\section{Sufficient conditions for detection}\label{subsec.detection}
We consider detection of alternatives $H_{0,j}^c$ or
$H_{0,G}^c$ with $|G| > 1$. We use again the notation $S_0$ as in Section
\ref{sec.multtest} and denote by $a_n \gg b_n$ that $a_n/b_n \to
\infty\ (n
\to\infty)$.
%
\begin{theo}\label{th.detection}
Consider the setting and assumptions as in Theorem~\ref{th1}.
\begin{enumerate}
\item
When considering individual hypotheses $H_{0,j}$: if $j \in S_0$ with
\[
\bigl|\beta^0_j\bigr| \gg a_{n,p;j}(\sigma)^{-1}\bigl|(P_{\bx})_{jj}\bigr|^{-1}
\max (\Delta_j,1)
\]
there exists an $\alpha_n \to0\ (n \to\infty)$ such that
\[
\PP[P_j \le\alpha_n] \to1\ (n \to\infty),
\]
while we still have for $j \in S_0^c$: $\limsup_{n\to\infty} \PP
[P_j \le
\alpha_n] - \alpha_n \le0$ (see Corollary~\ref{corr.pvalue}).
\item When considering individual hypotheses $H_{0,G}$ with $G = G_n$ and
$|G_n| > 1$: if
$H_{0,G}^c$ holds, with
\begin{eqnarray*}
\max_{j \in G_n} \bigl|a_{n,p;j}(\sigma) (P_{\bx})_{jj}^{-1}
\beta_j^0\bigr| \gg\max\Bigl(\max_{j \in G_n} |
\Delta_j|, \sqrt{\log\bigl(|G_n|\bigr)}\Bigr),
\end{eqnarray*}
there exists an $\alpha_n \to0\ (n \to\infty)$ such that
\[
\PP[P_{G_n} \le\alpha_n] \to1\ (n \to\infty),
\]
while if $H_{0,G}$ holds, $\limsup_{n\to\infty} \PP[P_{G_n}\le
\alpha_n] - \alpha_n \le0$ (see Corollary~\ref{corr.pvalue}).
\item When considering multiple hypotheses $H_{0,j}$: if for all $j \in
S_0$,
\[
\bigl|\beta^0_j\bigr| \gg a_{n,p;j}(\sigma)^{-1}\bigl|(P_{\bx})_{jj}\bigr|^{-1}
\max\bigl(\Delta_j,\sqrt{\log(p_n)}\bigr)
\]
there exists an $\alpha_n \to0\ (n \to\infty)$ such that
\begin{eqnarray*}
\PP[P_{\mathrm{corr};j} \le\alpha_n] \to1\ (n \to\infty)\ \mbox {for $j
\in S_0$}
\end{eqnarray*}
while we still have that $\limsup_{n \to\infty} \PP[V_{\alpha_n} >
0] -
\alpha_n \le0$ (see Theorem~\ref{th2}).
\item If in addition, $a_{n,p;j}(\sigma) \to\infty$ for all $j$
appearing in the conditions on $\beta_j^0$, we can replace in all the
statements 1--3 the ``$\gg$''
relation by ``$\ge\! C$'', where $0 < C < \infty$ is a sufficiently large
constant.
\end{enumerate}
\end{theo}

A proof is given in Section~\ref{sec.proofs}. Under the additional
assumption of Lemma~\ref{lemm.bound}, where the Lasso is used as initial
estimator and using the bounds in (\ref{bound1}), we obtain the bound (for
statement 1 in Theorem~\ref{th.detection}):
%
\begin{eqnarray}
\label{detection2a} \bigl|\beta^0_j\bigr| \ge C \max \biggl(
\frac{\max_{k \neq j}|(P_{\bx
})_{jk}|}{|(P_{\bx})_{jj}|} \biggl(\frac{\log(p_n)}{n} \biggr)^{1/2 - \xi},
\frac{1}{|(P_{\bx})_{jj}|} a_{n,p;j}(\sigma)^{-1} \biggr),
\end{eqnarray}
where $0 < \xi< 1/2$. This can be sharpened using the oracle bound,
assuming known order of sparsity:
\begin{eqnarray*}
\Delta_{\mathrm{orac;j}} = D s_{0,n} \max_{k \neq j} a_{n,p;j}(
\sigma) \bigl|(P_{\bx})_{jk}\bigr| \sqrt{\log(p_n)/n}
\end{eqnarray*}
for some $D>0$ sufficiently large (for example, assuming $s_{0,n}$ is
bounded, and replacing $s_{0,n}$ by~$1$ and choosing $D >0$ sufficiently
large). It then suffices to require
%
\begin{eqnarray}
\label{detection3} & &\bigl|\beta^0_j\bigr| \ge C\max \biggl(
\frac{\max_{k \neq j}|(P_{\bx
})_{jk}|}{|(P_{\bx})_{jj}|} s_{0,n} \biggl(\frac{\log(p_n)}{n}
\biggr)^{1/2},\frac{1}{|(P_{\bx
})_{jj}| a_{n,p;j}(\sigma)} \biggr)  \mbox{ for 1. in Th.
\ref{th.detection}},
\nonumber
\\[-8pt]
\\[-8pt]
\nonumber
& &\bigl|\beta^0_j\bigl| \ge C \max \biggl(\frac{\max_{k \neq j}|(P_{\bx
})_{jk}|}{|(P_{\bx})_{jj}|}
s_{0,n} \biggl(\frac{\log(p_n)}{n} \biggr)^{1/2},
\frac{\sqrt{\log(p_n)}}{|(P_{\bx})_{jj}|
a_{n,p;j}(\sigma)} \biggr) \mbox{ for 3. in Th.~\ref{th.detection}},\qquad\quad
\end{eqnarray}
and analogously for the second statement in Theorem
\ref{th.detection}.

\subsection{Order of magnitude of normalizing factors}\label{subsec.anrate}

The order of $a_{n,p;j}(\sigma)$ is typically much larger than $\sqrt{n}$
since in high dimensions, $\Omega_{jj}$ is very small. This means that the
Ridge estimator $\hat{\beta}_j$ has a much faster convergence rate than
$1/\sqrt{n}$ for estimating the projected parameter $\theta^0_j$.
This looks
counter-intuitive at first sight: the reason for the phenomenon is that
$\|\theta^0\|_2$ can be much smaller than $\|\beta^0\|_2$ and hence, Ridge
regression (which estimates the parameter $\theta^0$) is operating on
a much
smaller scale. This fact is
essentially an implication of the first statement in Lemma~\ref{lemm1}
(without the ``$\min_j$'' part). We can write
\[
\Omega_{jj} = \sum_{r=1}^n
\frac{s_r^2}{(s_r^2 + \lambda)^2} V_{jr}^2 = \sum
_{r=p-n+1}^p \frac{s_{r-p+n}^2}{(s_{r-p+n}^2 + \lambda)^2} U_{jr}^2,
\]
where the columns of $U = [U_{jr}]_{j,r=1,\ldots,p}$ contain the $p$
eigenvectors of $\bx^T\bx$, satisfying $\sum_{j=1}^p U_{jr}^2 = 1$.
For $n
\ll p$, only very few, namely $n$ terms, are left in the summation
while the
normalization for $U_{jr}^2$ is over all $p$ terms. For further
discussion about the fast convergence rate $a_{n,p;j}(\sigma)^{-1}$, see
Section~\ref{subsec.outlookbound}.\vadjust{\goodbreak}

While $a_{n,p;j}(\sigma)^{-1}$ is usually small, there is compensation with
$(P_{\bx})_{jj}^{-1}$ which can be rather large. In the detection
bound in
e.g., the first part of (\ref{detection3}), both terms appearing in the
maximum are often of the same order of magnitude; see also Figure
\ref{fig-supp1} in Section~\ref{subsec.outlookbound}. Assuming such a balance of terms, we obtain in
e.g., the first part of (\ref{detection3}):
\begin{eqnarray*}
\bigl|\beta^0_j\bigr| \ge C \frac{\max_{k \neq j}
|(P_{\bx})_{jk}|}{|(P_{\bx})_{jj}|} s_{0,n}
\sqrt{\log(p_n)/n}.
\end{eqnarray*}
The value of $\kappa_j = \max_{k \neq j}
|(P_{\bx})_{jk}|/|(P_{\bx})_{jj}|$ is often a rather small number between
0.05 and~4, see Table~\ref{tab1} in Section~\ref{sec.numeric}.
For comparison, \citet{zhangzhang11} establish under some conditions
detection for single hypotheses $H_{0,j}$ with $\beta_j^0$ in the
$1/\sqrt{n}$ range.
For the extreme case with $G_n = \{1,\ldots,p_n\}$, we are in
the setting of detection of the global hypotheses, see for example
\citet{ingsteretal10} for characterizing the detection boundary in
case of
independent covariables. Here, our analysis of detection is only providing
sufficient conditions, for rather general (fixed) design matrices.

\section{Numerical results}\label{sec.numeric}

As initial estimator for $\hat{\beta}_{\mathrm{corr}}$ in
(\ref{Ridgecorr}), we use the Scaled
Lasso with scale independent regularization parameter
$\lambda_{\mathrm{Scaled\mbox{-}Lasso}} = 2 \sqrt{\log(p)/n}$: it
provides an initial
estimate $\hat{\beta}_{\mathrm{init}}$ as well as an estimate
$\hat{\sigma}$ for the standard deviation $\sigma$. The parameter
$\lambda$ for Ridge regression in (\ref{Ridgetheta}) is always chosen as
$\lambda= 1/n$, reflecting the assumption in Theorem~\ref{th1} that it
should be small.

For single testing, we construct $p$-values as
in (\ref{pvalue1}) or (\ref{pvalue2}) with $\Delta_j$ from (\ref{bound1})
with $\xi= 0.05$. For multiple testing with familywise error control, we
consider $p$-values as in (\ref{pcorr}) with $\zeta=0$ (and $\Delta_j$ as
above).

\subsection{Simulations}\label{subsec.simul}

We simulate from the linear model as in (\ref{mod.lin}) with $\eps
\sim
\mathcal{ N}_n(0,I)$, $n = 100$ and the following configurations:
\begin{description}
\item[(M1)] For both $p \in\{500,2500\}$, the fixed design matrix is
generated from a
realization of $n$ i.i.d. rows from $\mathcal{ N}_p(0,I)$. Regarding the
regression coefficients, we consider active sets $S_0 =
\{1,2,\ldots,s_0\}$ with $s_0 \in\{3,15\}$ and three different
strengths of regression coefficients where $\beta^0_j \equiv b\ (j \in
S_0)$ with $b \in\{0.25,0.5,1\}$.\vspace*{1pt}
\item[(M2)] The same as in (M1) but for both $p \in\{500,2500\}$, the fixed
design matrix is
generated from a
realization of $n$ i.i.d. rows from $\mathcal{ N}_p(0,\Sigma)$ with
$\Sigma_{jk} \equiv0.8\ (j \neq k)$ and $\Sigma_{jj} = 1$.
\end{description}
The resulting signal to noise ratios $\mathrm{SNR} = \|\bx
\beta^0\|_2/\sigma$ are rather small:\vspace*{5pt}
\begin{center}
\begin{tabular}{@{}l|cccccc@{}}
$p\in\{500,2500\}$ & $(3,0.25)$ & $(3,0.5)$ & $(3,1)$ & $(15,0.25)$ &
$(15,0.5)$ &
$(15,1)$\\
\hline
(M1) &0.46 & 0.93 & 1.86 & 1.06 & 2.13 & \phantom{0}4.26\\
(M2) &0.65& 1.31 & 2.62 & 3.18 & 6.37 & 12.73
\end{tabular}
\end{center}\eject
Here, a pair such as $(3,0.25)$ denotes the values of $s_0=3,\ b=0.25$
(where $b$ is the value of the active regression coefficients).

\begin{figure}[t]
\centering
\begin{tabular}{@{}cc@{}}

\includegraphics{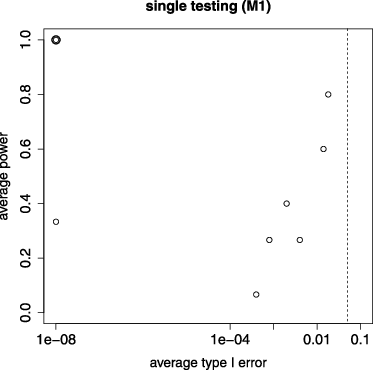}
 & \includegraphics{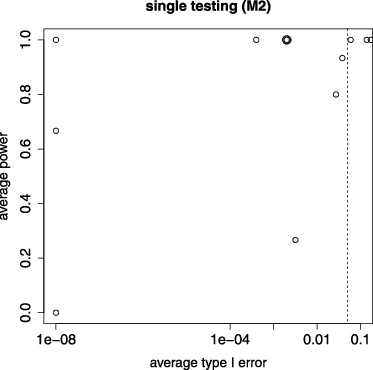}\\
\footnotesize{(a)} & \footnotesize{(b)}\\[3pt]

\includegraphics{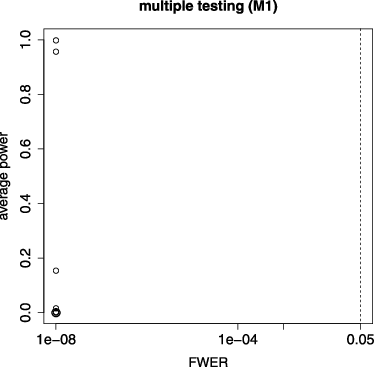}
 & \includegraphics{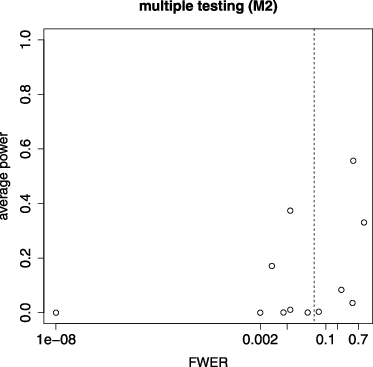}\\
\footnotesize{(c)} & \footnotesize{(d)}
\end{tabular}
\caption{Simulated data as described in Section~\protect\ref
{subsec.simul}. \textup{(a)} and
\textup{(b)}: Single testing with average
type I error~(\protect\ref{avetypeI}) on x-axis (log-scale) and
average power
(\protect\ref{avepower}) on y-axis. \textup{(c) }and \textup{(d)}: Multiple testing
with familywise
error rate on x-axis (log-scale) and average power (\protect\ref
{avepower}), but
using $P_{\mathrm{corr};j}$, on y-axis. Vertical dotted line is at
abscissa $0.05$. Each point corresponds to a model
configuration. \textup{(a)} and \textup{(c)}: 12 model configurations generated from independent
covariates \textup{(M1)}; \textup{(b)} and \textup{(d)}: 12 model configurations generated from
equi-dependent covariates \textup{(M2)}. When an error is zero, we plot it on the
log-scale at abscissa $10^{-8}$.} \label{fig1}\vspace*{-6pt}
\end{figure}

We consider the decision-rule at significance level $\alpha= 0.05$
%
\begin{equation}
\label{desc-rule} \mbox{reject $H_{0,j}$ if $P_j \le0.05$},
\end{equation}
for testing single hypotheses where $P_j$ is as in (\ref{pvalue1}) with
plugged-in estimate $\hat{\sigma}$. The considered type I error is the
average over non-active variables:
%
\begin{equation}
\label{avetypeI} (p - s_0)^{-1} \sum
_{j \in S_0^c} \PP[P_j \le0.05]
\end{equation}
and the average power is
%
\begin{equation}
\label{avepower} s_0^{-1} \sum_{j \in S_0}
\PP[P_j \le0.05].
\end{equation}
For multiple testing, we consider the adjusted $p$-value
$P_{\mathrm{corr};j}$ from (\ref{pcorr}):
the decision is as in~(\ref{desc-rule}) but replacing $P_j$ by
$P_{\mathrm{corr};j}$. We
report the familywise
error rate (FWER) $\PP[V_{0.05} > 0]$ and the average power as in
(\ref{avepower}) but the latter with using $P_{\mathrm{corr};j}$.
The results are displayed in Figure~\ref{fig1}, based on 500 simulation
runs per setting (with the same fixed design per setting).
The subfigure (d) shows that the proposed method exhibits essentially four
times a too large familywise error rate in multiple testing: it happens for
scenarios with strongly correlated variables (model (M2)) and where the
sparsity $s_0 = 15$ is large with moderate or large size of the
coefficients (scenario (M2) with $s_0=15$ and coefficient size $b=0.25$ is
unproblematic). The corresponding number of false positives are
reported in
Table~\ref{tab-supp.1} in Section~\ref{sec.falsepos}.

\subsection{Values of $P_{\bx}$}
The detection results in (\ref{detection2a}) and (\ref{detection3}) depend
on the ratio $\kappa_j = \max_{k \neq
j}|(P_{\bx})_{jk}|/|(P_{\bx})_{jj}|$. We report in Table~\ref{tab1} summary
statistics of $\{\kappa_j\}_j$ for various datasets.
We clearly see that the values of $\kappa_j$ are typically rather small
which implies good detection properties as discussed in Section~\ref{subsec.detection}. Furthermore, the values $\max_{k \neq
j}|(P_{\bx})_{jk}|$
occurring in the construction of $\Delta_j$ in Section~\ref{subsec.Delta}
are typically very small (not shown here).

%
\begin{table}
\def\arraystretch{0.9}
\caption{Minimum, maximum and three quartiles of $\{\kappa_j\}_{j=1}^p$ for
various designs $\bx$ from different datasets. The first four are from the
simulation models in Section \protect\ref{subsec.simul}. Although not relevant
for the table, ``Motif'' (see Section \protect\ref{subsec.realdata}) and
``Riboflavin'' have a continuous response while the last six have a class
label \protect\citep{dett04}}\label{tab1}\vspace*{-3pt}
\begin{tabular*}{\textwidth}{@{\extracolsep{\fill}}llllll@{}}
\hline
dataset, $(n,p)$ & $\min_j \kappa_j$ & $0.25$q$\{\kappa_j\}_j$
& med$\{\kappa_j\}_j$ & $0.75$q$\{\kappa_j\}_j$ & $\max_j \kappa_j$ \\
\hline
(M1), $(100,500)$ & 0.21 & 0.27 & 0.29 & 0.31 & 0.44 \\
(M1), $(100,2500)$ & 0.27 & 0.34 & 0.36 & 0.39 & 0.54 \\
(M2), $(100,500)$ & 0.20 & 0.26 & 0.29 & 0.32 & 0.45 \\
(M2), $(100,2500)$ & 0.26 & 0.33 & 0.36 & 0.39 & 0.59 \\
Motif, $(143,287)$ & 0.05 & 0.10 & 0.13 & 0.18 & 0.47 \\
Riboflavin, $(71,4088)$ & 0.29 & 0.54 & 0.65 & 0.77 & 1.73\\
Leukemia, $(72,3571)$ & 0.32 & 0.44 & 0.50 & 0.58 & 1.57 \\
Colon, $(62,2000)$ & 0.28 & 0.50 & 0.57 & 0.67 & 1.36 \\
Lymphoma, $(62,4026)$ & 0.34 & 0.52 & 0.63 & 0.78 & 1.49 \\
Brain, $(34,5893)$ & 0.51 & 0.63 & 0.67 & 0.74 & 2.44 \\
Prostate, $(102,6033)$ & 0.26 & 0.45 & 0.57 & 0.74 & 3.67 \\
NCI, $(61,5244)$ & 0.37 & 0.52 & 0.61 & 0.79 & 1.76\\
\hline
\end{tabular*}
\end{table}
%

\subsection{Real data application}\label{subsec.realdata}

We consider a problem about motif regression for finding the
binding sites in DNA sequences of the HIF1$\alpha$ transcription
factor. The binding sites are also called motifs, and they are typically
6--15 base pairs (with categorical values $\in\{A,C,G,T\}$) long.

The data consists of a univariate response variable $Y$ from CHIP-chip
experiments, measuring the
logarithm of the binding intensity of the HIF1$\alpha$ transcription factor
on coarse DNA segments. Furthermore, for each DNA segment, we have
abundance scores for $p=195$ candidate motifs, based on
DNA sequence data. Thus, for each DNA segment\vadjust{\eject} $i$ we have $Y_i \in\R$ and
$X_i\in\R^p$, where $i=1,\ldots,n_{\mathrm{tot}}=287$ and $p=195$. We
consider a linear model as in (\ref{mod.lin}) and hypotheses
$H_{0,j}$ for $j=1,\ldots,p=195$: rejection of $H_{0,j}$ then corresponds
to a significant motif.
This dataset has been analyzed in \citet{memepb09} who found one
significant motif using their
$p$-value method for a linear model based on multiple sample splitting (which
assumes the unpleasant ``beta-min'' condition in (\ref{beta.min})).

\begin{figure}
\centering
\begin{tabular}{@{}cc@{}}

\includegraphics{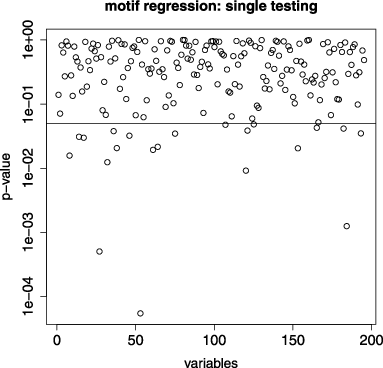}
 & \includegraphics{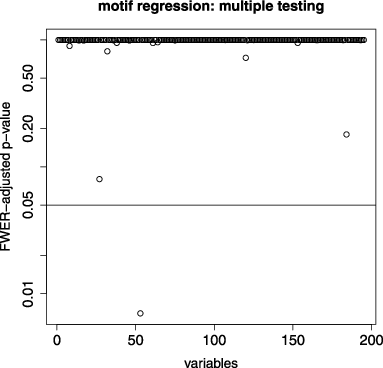}\\
\footnotesize{(a)} & \footnotesize{(b)}
\end{tabular}
\caption{Motif regression with $n=143$ and $p=195$. \textup{(a)} Single-testing
$p$-values as in (\protect\ref{pvalue1}); \textup{(b)} Adjusted $p$-values as in
(\protect\ref{pcorr})
for FWER control. The $p$-values are plotted on the log-scale. Horizontal
line is at $y=0.05$.}
\label{fig3}
\end{figure}

Since the dataset has $n_{\mathrm{tot}} > p$ observations, we take one
random subsample of size $n = 143 < p=195$. Figure~\ref{fig3} reports the
single-testing as well as the adjusted $p$-values for controlling the
FWER. There
is one significant motif with corresponding FWER-adjusted $p$-value equal
to 0.007, and the method in \citet{memepb09} based on
the total sample with $n_{\mathrm{tot}}$ found the same significant
variable with FWER-adjusted $p$-value equal to 0.006. Interestingly, the
weakly significant motif with $p$-value 0.080 is known to be a true
binding site for HIF1$\alpha$, thanks to biological validation experiments.

When compared to the Bonferroni--Holm
procedure for controlling FWER based on the raw $p$-values as shown in Figure
\ref{fig3}(a), we have for the variables with smallest $p$-values:
\begin{eqnarray*}
\mbox{method as in (\ref{pcorr}):}& &\ 0.007,\ 0.080,\ 0.180,
\\
\mbox{Bonferroni--Holm:}& &\ 0.011,\ 0.098,\ 0.242.
\end{eqnarray*}
Thus, for this example, the multiple testing correction as in Section
\ref{sec.multtest} does not provide large improvements in power over the
Bonferroni--Holm procedure; but our method is closely related to the
Westfall--Young procedure which has been shown to be asymptotically
optimal for a
broad class of high-dimensional problems (Meinshausen, Maathuis, and B\"uhlmann,
\citeyear{memabu11}).

\section{Finite sample results}\label{sec.finites}

We present here finite sample analogues of Theorem~\ref{th1} and
\ref{th2}. Instead of assumption \textup{(A)}, we assume the
following:\vadjust{\eject}
\begin{description}
\item[(A$^{\prime}$)] There are constants $\Delta_j > 0$ such that
\begin{eqnarray*}
\PP\biggl[\bigcap_{j=1} \biggl\{a_{n,p;j}(\sigma) \sum
_{k \neq j} (P_{\bx
})_{jk}\bigl(\hat{
\beta}_{\mathrm{init};k} - \beta^0_k\bigr)| \le
\Delta_j\biggr\}\biggr] \ge1 - \kappa
\end{eqnarray*}
for some (small) $0 < \kappa< 1$.
\end{description}
We then have the following result.
%
\begin{prop}\label{prop2}
Assume model (\ref{mod.lin}) with Gaussian errors. Consider the corrected
Ridge regression estimator $\hat{\beta}_{\mathrm{corr}}$ in
(\ref{Ridgecorr}) with regularization parameter $\lambda> 0$,
and assume (\ref{minvar}) and condition~\textup{(A$^{\prime}$)}. Then, with probability
at least
$1 - \kappa$, for $j \in\{1,\ldots,p\}$ and if $H_{0,j}$ holds:
\begin{eqnarray*}
& &a_{n,p;j}(\sigma) |\hat{\beta}_{\mathrm{corr};j}| \le a_{n,p;j}(
\sigma) |Z_j| + \Delta_j + \bigl\|a_{n,p} b(\lambda)
\bigr\|_{\infty},
\\
& &\bigl\|a_{n,p} b(\lambda)\bigr\|_{\infty} = \max_{j=1,\ldots,p}
a_{n,p;j}(\sigma) \bigl|b_j(\lambda)\bigr| \le \frac{\lambda}{\Omega_{\mathrm{min}}(\lambda
)^{1/2}}
n^{1/2} \sigma^{-1} \bigl\|\theta^0\bigr\|_2
\lambda_{\mathrm{min} \neq0}(\hat {\Sigma})^{-1}.
\end{eqnarray*}
Similarly, with probability at least
$1 - \kappa$, for any subset $G \subseteq\{1,\ldots,p\}$ and if $H_{0,G}$
holds:
\begin{eqnarray*}
& &\max_{j \in G} a_{n,p;j}(\sigma) |\hat{\beta}_{\mathrm
{corr};j}| \le
\max_{j \in
G} \bigl(a_{n,p;j}(\sigma) |Z_j| +
\Delta_j \bigr) + \bigl\|a_{n,p} b(\lambda)\bigr\|_{\infty}.
\end{eqnarray*}
\end{prop}

A proof is given in Section~\ref{sec.proofs}.
Due to the third statement in Lemma~\ref{lemm1},
$\Omega_{\mathrm{min}}(\lambda)^{-1/2}$ is bounded for a bounded
range of
$\lambda\in(0,C]$. Therefore, the bound for $\|a_{n,p}
b(\lambda)\|_{\infty}$ can be made arbitrarily small by choosing
$\lambda>
0$ sufficiently small.

Theorem~\ref{th2} is a consequence of the following finite sample result.
%
\begin{prop}\label{prop3}
Consider the event $\mathcal{ E}$ with probability $\PP[\mathcal{
E}] \ge1 -
\kappa$ where condition \textup{(A$^{\prime}$)} holds. Then, when using the corrected $p$-values
from (\ref{pcorr}), with $\zeta\ge
0$ (allowing also $\zeta=0$), we obtain
approximate strong control of the familywise error rate:
\begin{eqnarray*}
\PP[V_{\alpha}>0] \le F_Z\bigl(F_Z^{-1}(
\alpha) - \zeta+ 2 (2\pi)^{-1/2} \bigl\|a_{n,p} b(\lambda)
\bigr\|_{\infty}\bigr) + \bigl(1 - \PP[\mathcal{ E}]\bigr).
\end{eqnarray*}
\end{prop}

A proof is given in Section~\ref{sec.proofs}. We immediately get the following
bound for $\zeta\ge0$:
\begin{eqnarray*}
\PP[V_{\alpha}>0] \le\alpha+ \sup_u\bigl|F'_Z(u)\bigr|
2(2\pi)^{-1/2}\bigl \|a_{n,p} b(\lambda)\bigr\|_{\infty} + \bigl(1
- \PP[\mathcal{ E}]\bigr).
\end{eqnarray*}

\section{Conclusions}
We have proposed a novel construction of $p$-values for individual and more
general hypotheses in a high-dimensional linear model with fixed design and
Gaussian errors. We have restricted ourselves to max-type statistics for
general hypotheses but modifications to e.g., weighted sums are
straightforward using the representation in Proposition \ref
{prop-repr}. A~key idea is to use a linear, namely the Ridge
estimator, combined with a correction for the potentially substantial bias
due to the fact that the Ridge estimator is estimating the projected
regression parameter vector onto the row-space of the design matrix. The
finding that we
can ``succeed'' with a corrected Ridge estimator in a high-dimensional context
may come as a surprise, as it is well known that Ridge estimation can be
very bad for say prediction. Nevertheless, our bias corrected Ridge
procedure might not be optimal in terms of power, as indicated in Section
\ref{subsec.anrate}. The main
assumptions we make are the compatibility condition for the design, i.e.,
an identifiability condition, and knowledge of an upper bound of
the sparsity (see Lemma~\ref{lemm.bound}).
A related idea of using
a linear estimator coupled with a bias correction for deriving
confidence intervals has been earlier proposed by
\citet{zhangzhang11}.

\emph{No tuning parameter.}
Our approach does not require the specification of a tuning
parameter, except for the issue that we crudely bound the true sparsity
as in
(\ref{bound1}); we always used $\xi= 0.05$, and the Scaled Lasso initial
estimator does not
require the specification of a regularization parameter. All our numerical
examples were run without tuning the method to a specific setting, and
error control with our $p$-value approach is often conservative while the
power seems reasonable.
Furthermore, our method is
generic which allows to test for any $H_{0,G}$ regardless whether the size
of $G$ is small or large: we present in the Section
\ref{sec.addsimul} an additional simulation where $|G|$ is large. For
multiple testing correction or for general
hypotheses with sets $G$ where $|G| > 1$, we rely on the power of
simulation since analytical formulae for max-type statistics under
dependence seem in-existing: yet, our simulation is extremely simple as we
only need to generate dependent multivariate Gaussian random
variables.

\emph{Small variance of Ridge estimator.}
As indicated before, it is surprising that corrected Ridge estimation performs
rather well for statistical testing. Although the bias due
to the projection $P_{\bx}$ can be substantial, it is compensated by small
variances $\sigma^2 n^{-1} \Omega_{jj}$ of the
Ridge estimator. It is \emph{not} true that $\Omega_{jj}$'s become large
as $p$ increases: that is, the Ridge estimator has small variance for an
individual component when $p$ is very large, see Section
\ref{subsec.anrate}. Therefore, the detection power of the method remains
reasonably good as discussed in Section
\ref{subsec.detection}. Viewed from a different perspective, even though
$|(P_{\bx})_{jj} \beta^0_j|$ may be very small, the normalized version
$a_{n,p;j}(\sigma) |(P_{\bx})_{jj} \beta^0_j|$ can be sufficiently
large for
detection since $a_{n,p;j}(\sigma)$ may be very large (as the inverse of
the square root of the variance). The values of $P_{\bx}$ can be easily
computed for a given problem: our analysis about sufficient conditions for
detection in Section
\ref{subsec.detection} could be made more complete by invoking random matrix
theory for the projection $P_{\bx}$ (assuming that $\bx$ is a
realization of
i.i.d. row-vectors whose entries are potentially dependent). However,
currently, most of the results on singular values
and similar quantities of $\bx$ are for the regime $p \le n$
\citep{versh12}, which leads in our context to the trivial projection
$P_{\bx} = I$, or for the regime $p/n \to C$ with $0\le C <\infty$
\citep{elkar08}.

\emph{Extensions.}
Obvious but partially non-trivial model extensions include random design,
non-Gaussian errors or generalized linear models. From a
practical point of view, the second and third issue would be most
valuable. Relaxing the fixed design assumption makes part of the
mathematical arguments more complicated, yet a random design is better
posed in terms of identifiability.\vspace*{-3pt}

\begin{appendix}
\section*{Appendix}\label{sec.supplement}\vspace*{-3pt}

\subsection{Proofs}\label{sec.proofs}\vspace*{-3pt}

\begin{pf*}{Proof of Proposition~\ref{prop1}}
The statement about the bias is given in \citet{shadeng11} (proof of their
Theorem 1). The covariance matrix of $\hat{\beta}$ is
\[
n^{-1} \Omega= n^{-1} (\hat{\Sigma} + \lambda I)^{-1}
\hat{\Sigma} (\hat{\Sigma} + \lambda I)^{-1}.
\]
Then, for the variance we obtain $\Var(\hat{\beta}_j) = n^{-1}
\sigma^2
\Omega_{jj} \ge n^{-1} \sigma^2 \Omega_{\mathrm{min}}(\lambda
)$.
\end{pf*}

\begin{pf*}{Proof of Proposition~\ref{prop-repr}}
We write
\begin{eqnarray*}
\hat{\beta}_{\mathrm{corr};j} = \bigl(\hat{\beta}_j - \EE[\hat{\beta
}_j]\bigr) + \theta^0_j - \sum
_{k \neq j} (P_{\bx})_{jk} \hat{
\beta}_{\mathrm
{init};k} + \bigl(\EE[\hat{\beta}_j] -
\theta^0_j\bigr).
\end{eqnarray*}
The result then follows by defining $Z_j = \hat{\beta}_j -
\EE[\hat{\beta}_j]$ and using that $\theta^0_j = (P_{\bx} \beta^0)_j =
(P_{\bx})_{jj} \beta^0_j + \sum_{k \neq j} (P_{\bx})_{jk}
\beta^0_k$.
\end{pf*}

\begin{pf*}{Proof of Proposition~\ref{prop2} (basis for proving Theorem
\ref{th1})}
The bound from Proposition~\ref{prop1} for the estimation bias of the Ridge
estimator leads to:
\begin{eqnarray*}
\bigl\|a_{n,p} b(\lambda)\bigl\|_{\infty} &=& \max_{j=1,\ldots,p}
a_{n,p;j}(\sigma) \bigl|\EE[\hat{\beta}_j] - \theta^0_j\bigr|
\\
&\le& \frac{\lambda\|\theta^0\|_2 \lambda_{\mathrm{min} \neq
0}(\hat{\Sigma})^{-1}}{\sigma n^{-1/2} \Omega_{jj}^{1/2}}
\\
&\le&\lambda\bigl\|\theta^0\bigl\|_2 \lambda_{\mathrm{min} \neq
0}(\hat{
\Sigma})^{-1} \sigma^{-1} n^{1/2} \Omega_{\mathrm
{min}}(
\lambda)^{-1/2} .
\end{eqnarray*}
By using the representation from Proposition~\ref{prop-repr}, invoking
assumption \textup{(A$^{\prime}$)} and assuming that the null-hypothesis $H_{0,j}$ or
$H_{0,G}$ holds, respectively, the proof is completed.
\end{pf*}

\begin{pf*}{Proof of Theorem~\ref{th1}}
Due to the choice of $\lambda= \lambda_n$ we have that $\|a_{n,p}
b(\lambda_n)\|_{\infty} = o(1)\ (n \to\infty)$. The proof then
follows from
Proposition~\ref{prop2} and invoking assumption \textup{(A)} saying that the
probabilities for the statements in Proposition~\ref{prop2} converge
to 1
as $n \to\infty$.
\end{pf*}

\begin{pf*}{Proof of Proposition~\ref{prop3} (basis for proving Theorem
\ref{th2})}
Consider the set $\mathcal{ E}$ where assumption \textup{(A$^{\prime}$)} holds (whose probability
is at least $\PP[\mathcal{ E}] \ge1- \kappa$). Without loss of
generality, we
consider $P_j = 2 (1 - \Phi(a_{n,p;j}(\sigma)|\hat{\beta
}_{\mathrm{corr};j}| -
\Delta_j))$ without the truncation at value 1 (implied by the positive part
$(a_{n,p;j}(\sigma)|\hat{\beta}_{\mathrm{corr};j}| - \Delta_j)_+$); in
terms of decisions (rejection or non-rejection of a hypothesis), both
versions for the $p$-value are equivalent.
Then, on $\mathcal{ E}$ and for $j
\in S_0^c$:
\begin{eqnarray*}
P_j &=& 2 \bigl(1 - \Phi\bigl(a_{n,p;j}(\sigma)|\hat{
\beta}_{\mathrm
{corr};j}| - \Delta_j\bigr) \bigr)
\\
&\ge& 2 \biggl(1 - \Phi\biggl(a_{n,p;j}(\sigma) \biggl|\hat{\beta
}_{\mathrm{corr};j} - \sum_{k \neq j} (P_{\bx})_{jk}
\bigl(\hat{\beta}_{\mathrm{init};k} - \beta^0_k\bigr)\biggr |
\biggr) \biggr)
\\
&\ge& 2 \bigl(1 - \Phi\bigl(a_{n,p;j}(\sigma)|Z_j|\bigr)
\bigr) - 2 (2\pi)^{-1/2} \bigl\|a_{n,p} b(\lambda)\bigr\|_{\infty},
\end{eqnarray*}
where in the last inequality we used Proposition~\ref{prop-repr} and Taylor's
expansion.
Thus, on $\mathcal{ E}$:
\begin{eqnarray*}
\min_{j \in S_0^c} P_j &\ge& \min_{j \in S_0^c} 2 \bigl(1 - \Phi
\bigl(a_{n,p;j}(\sigma)|Z_j|\bigr) \bigr) - 2(2
\pi)^{-1/2} \bigl\|a_{n,p} b(\lambda)\bigr\|_{\infty}
\\
&\ge& \min_{j=1,\ldots,p} 2 \bigl(1 - \Phi\bigl(a_{n,p;j}(
\sigma)|Z_j|\bigr) \bigr) - 2(2 \pi)^{-1/2}
\bigl\|a_{n,p} b(\lambda)\bigr\|_{\infty}.
\end{eqnarray*}
Therefore,
\begin{eqnarray*}
\PP\Bigl[\min_{j \in S_0^c} P_j \le c\Bigr]& \le&\PP\Bigl[
\mathcal{ E} \cap \Bigl\{\min_{j \in
S_0^c} P_j \le c\Bigr\}\Bigr] + \PP\bigl[
\mathcal{ E}^c\bigr]
\\
&\le& \PP\Bigl[\min_{j=1,\ldots,p} 2 \bigl(1 - \Phi\bigl(a_{n,p;j}(
\sigma )|Z_j|\bigr) \bigr) \le c + 2(2 \pi)^{-1/2}
\bigl\|a_{n,p}b(\lambda)\bigr\|_{\infty}\Bigr] + \PP\bigl[\mathcal{
E}^c\bigr]
\\
&=& F_Z \bigl(c + 2(2 \pi)^{-1/2} \bigl\|a_{n,p}b(
\lambda)\bigr\|_{\infty} \bigr) + \PP\bigl[\mathcal{ E}^c\bigr].
\end{eqnarray*}
Using this we obtain:
\begin{eqnarray*}
\PP[V_{\alpha}>0] &=& \PP\Bigl[\min_{j \in S_0^c} P_{\mathrm{corr};j} \le
\alpha\Bigr] = \PP\Bigl[\min_{j \in S_0^c} P_j \le
F_Z^{-1}(\alpha) - \zeta\Bigr]
\\
&\le& F_Z \bigl(F_Z^{-1}(\alpha) - \zeta+ 2
(2\pi)^{-1/2} \bigl\|a_{n,p} b(\lambda)\bigr\|_{\infty} \bigr) +
\PP\bigl[\mathcal{ E}^c\bigr].
\end{eqnarray*}
This completes the proof.
\end{pf*}

\begin{pf*}{Proof of Theorem~\ref{th2}}
Due to the choice of
$\lambda= \lambda_n$ we have that $\|a_{n,p}
b(\lambda_n)\|_{\infty} = o(1)\ (n \to\infty)$. Furthermore, using the
formulation in Proposition~\ref{prop3}, assumption \textup{(A)}
translates to a sequence of sets $\mathcal{ E}_n$ with $\PP[\mathcal
{ E}_n] \to
1\ (n \to\infty)$. We then use Proposition~\ref{prop3} and observe
that for
sufficiently large $n$:
$F_Z(F_Z^{-1}(\alpha) - \zeta+ 2(2 \pi)^{-1/2} \|a_{n,p} b(\lambda_n)\|_{\infty}) \le
F_Z(F_Z^{-1}(\alpha)) \le\alpha$. The modification for the case with
$\alpha_n \to0$ sufficiently slowly follows analogously: note that the
second last inequality in the proof above follows by monotonicity of
$F_Z(\cdot)$ and $\zeta> 2 (2 \pi)^{-1/2} \|a_{n,p} b(\lambda_n)\|_{\infty}$
for $n$ sufficiently large. This completes the proof.
\end{pf*}

\begin{pf*}{Proof of Theorem~\ref{th.detection}}
Throughout the proof, $\alpha_n \to0$ is converging sufficiently slowly,
possibly depending on the context of the different statements we prove.
Regarding statement 1: it is sufficient that for $j \in S_0$,
\[
a_{n,p;j}(\sigma) |\hat{\beta}_{\mathrm{corr};j}| \gg\max(
\Delta_j,1).
\]
From Proposition~\ref{prop-repr}, we see that this can be enforced by requiring
\begin{eqnarray*}
a_{n,p;j}(\sigma) \biggl(\bigl|(P_{\bx})_{jj}
\beta^0_j\bigr| - \biggl|\sum_{k \neq
j}
(P_{\bx})_{jk} \bigl(\hat{\beta}_{\mathrm{init};k} -
\beta^0_k\bigr)\biggr| - |Z_j| -
\bigl|b_j(\lambda )\bigr| \biggr) \gg \max(\Delta_j,1).
\end{eqnarray*}
Since $|a_{n,p;j}(\sigma)\sum_{k \neq j} (P_{\bx})_{jk}
(\hat{\beta}_{\mathrm{init};k} - \beta^0_k)| \le\Delta_j$, this
holds if
%
\begin{eqnarray}
\label{add-detect1}\bigl |\beta^0_j\bigr| \gg\frac{1}{|(P_{\bx})_{jj}| a_{n,p;j}(\sigma)}
\max\bigl(\Delta_j,a_{n,p;j}(\sigma) Z_j,
a_{n,p;j}(\sigma) b_j(\lambda),1\bigr).
\end{eqnarray}
Due to the choice of $\lambda= \lambda_n$ (as in Theorem~\ref{th1}), we
have $a_{n,p;j}(\sigma) b_j(\lambda) \le\|a_{n,p}(\sigma)
b(\lambda)\|_{\infty} = o(1)$. Hence, (\ref{add-detect1}) holds with
probability converging to one if
\[
\bigl|\beta^0_j\bigr| \gg\frac{1}{|(P_{\bx})_{jj}| a_{n,p;j}(\sigma)} \max(
\Delta_j,1),
\]
completing the proof for statement 1.

For proving the second statement, we recall that
\begin{eqnarray*}
1 - J_G(c) = \PP\Bigl[\max_{j \in G} \bigl(a_{n,p;j}(
\sigma) |Z_j| + \Delta_j \bigr) > c\Bigr].
\end{eqnarray*}
Denote by $W = \max_{j \in G} (a_{n,p;j}(\sigma) |Z_j| + \Delta_j)
\le
\tilde{W} = \max_{j \in G} a_{n,p;j}(\sigma) |Z_j| + \max_{j \in G}
\Delta_j$. Thus,
\[
\PP[W > c] \le\PP[\tilde{W} > c].
\]
Therefore, the statement for the $p$-value $\PP[P_G \le
\alpha_n]$ is implied by
%
\begin{equation}
\label{add-detect2} \PP_{\tilde{W}}[\tilde{W} > \hat{\gamma}_G]
\le\alpha_n.
\end{equation}
Using the union bound and the fact that $a_{n,p;j}(\sigma) |Z_j| \sim
\mathcal{
N}(0,1)$ (but dependent over different values of $j$), we have that
\[
\max_{j \in G} a_{n,p;j}(\sigma) |Z_j| =
O_P\bigl(\sqrt{\log\bigl(|G|\bigr)}\bigr).
\]
Therefore, (\ref{add-detect2}) holds if
\begin{eqnarray*}
\hat{\gamma}_G = \max_{j \in G} a_{n,p;j}(\sigma) |
\hat{\beta}_{\mathrm{corr};j}| \gg\max\Bigl(\max_{j \in G} \Delta_j,
\sqrt{\log\bigl(|G|\bigr)}\Bigr).
\end{eqnarray*}
The argument is now analogous to the proof of the first statement above,
using the representation from Proposition~\ref{prop-repr}.

Regarding the third statement, we invoke the rough bound
\[
P_{\mathrm{corr};j} \le p P_j,
\]
with the non-truncated Bonferroni corrected $p$-value at the right-hand
side. Hence,
\[
\max_{j \in S_0} P_{\mathrm{corr};j} \le\alpha_n
\]
is implied by
\begin{eqnarray*}
\max_{j \in S_0} pP_j = \max_{j \in S_0} 2p \bigl(1 - \Phi
\bigl(\bigl(a_{n,p;j}(\sigma) |\hat{\beta}_{\mathrm{corr};j}| -
\Delta_j\bigr)_+\bigr)\bigr) \le\alpha_n.
\end{eqnarray*}
Since this involves a standard Gaussian two-sided tail probability, the
inequality can be enforced (for certain slowly converging $\alpha_n$) by
\begin{eqnarray*}
\max_{j \in S_0} 2\exp \bigl(\log(p) - \bigl(a_{n,p;j}(\sigma) |\hat{
\beta}_{\mathrm{corr};j}| - \Delta_j\bigr)_+^2/2 \bigr) =
o_P(1).
\end{eqnarray*}
The argument is now analogous to the proof of the first statement above,
using the representation from Proposition~\ref{prop-repr}.

The fourth statement involves slight obvious
modifications of the arguments above.
\end{pf*}

\subsection{$P$-values for $H_{0,G}$ with $|G|$ large}\label{sec.addsimul}

We report here on a small simulation study for testing $H_{0,G}$ with
$G =
\{1,2,\ldots,100\}$. We consider model (M2) from Section
\ref{subsec.simul} with 4 different configurations and we use the $p$-value
from~(\ref{pvalue2}) with corresponding decision rule for rejection of
$H_{0,G}$ if the $p$-value is smaller or equal to the nominal level
0.05. Table~\ref{tab-supp2} describes the result based on 500 independent
simulations (where the fixed design remains the same).
The method works well with much better power\vadjust{\eject} than multiple testing of
individual hypotheses but worse than average power for testing
individual hypotheses without multiplicity adjustment (which is not a
proper approach). This is largely in
agreement with the theoretical results in Theorem
\ref{th.detection}. Furthermore, the
type I error control is good.

%
\renewcommand{\thetable}{\arabic{table}}
\setcounter{table}{1}
\begin{table}
\def\arraystretch{0.9}
\tabcolsep=4pt
\caption{Testing of general hypothesis $H_{0,G}$ with $|G| = 100$ using
the $p$-value in (\protect\ref{pvalue2}) with significance level
$0.05$. Second
column: type I error; Third column: power; Fourth column: comparison with
power using multiple individual testing and average power using individual
testing without multiplicity adjustment (both for all $p$ hypotheses
$H_{0,j}\
(j=1,\ldots,p)$)}\label{tab-supp2}
\begin{tabular*}{\textwidth}{@{\extracolsep{\fill}}lccc@{}}
\hline
Model & $\PP[\mbox{false rejection}]$ & $\PP[\mbox{true
rejection}]$ &
(power mult., power indiv.)\\
\hline
(M2), $p=500$, $s=3$, $b=0.5$ & 0.00 & 0.10 & (0.01,1.00)\\
(M2), $p=500$, $s=3$, $b=1$ & 0.00 & 0.91 & (0.37,1.00)\\
(M2), $p=2500$, $s=3$, $b=0.5$ & 0.01 & 0.02 & (0.00,1.00)\\
(M2), $p=2500$, $s=3$, $b=1$ & 0.00 & 0.83 & (0.17,1.00)
\\
\hline
\end{tabular*}\vspace*{-6pt}
\end{table}
%

\subsection{Number of false positives in simulated
examples}\label{sec.falsepos}
We show in Table~\ref{tab-supp.1} the number of false positives $V = V_{0.05}$ in the
simulated scenarios where the FWER (among individual hypotheses) was found
too large.
Although the FWER is larger than 0.05, the number of false
positives is relatively small, except for the extreme model
(M2), $p=2500$, $s=15$, $b=1$ which has a too large sparsity and a
too strong signal strength. For the latter model, we would need to increase
$\xi$ in (\ref{bound1}) to achieve better error control.

%
\begin{table}[b]\vspace*{-6pt}
\tabcolsep=0pt
\def\arraystretch{0.9}
\caption{Probabilities for false positives for simulation models from
Section \protect\ref{subsec.simul} in scenarios where the FWER is
clearly overshooting
the nominal level $0.05$}\label{tab-supp.1}
\begin{tabular*}{\textwidth}{@{\extracolsep{\fill}}lcccccc@{}}
\hline
Model & $\PP[V=0]$ & $\PP[V=1]$ & $\PP[V=2]$ & $\PP[V = 3]$ & $\PP
[V =4]$ &
$\PP[V \ge5]$ \\
\hline
(M2), $p=500$, $s=15$, $b=1$ & 0.482 & 0.336 & 0.138 & 0.028 & 0.010 &
0.006 \\
(M2), $p=500$, $s=15$, $b=0.5$ & 0.746 & 0.218 & 0.034 & 0.000 & 0.002 &
0.000 \\
(M2), $p=2500$, $s=15$, $b=1$ & 0.012 & 0.044 & 0.098 & 0.126 & 0.172 &
0.548 \\
(M2), $p=2500$, $s=15$, $b=0.5$ & 0.504 & 0.328 & 0.132 & 0.032 & 0.004 &
0.000\\
\hline
\end{tabular*}
\end{table}
%

\subsection{\texorpdfstring{Further discussion about $p$-values and bounds $\Delta_j$ in assumption \textup{(A)}}
{Further discussion about p-values and bounds Delta j in assumption (A)}}\label{subsec.outlookbound}

The $p$-values in (\ref{pvalue1}) and (\ref{pvalue2}) are crucially
based on
the idea of correction with the bounds $\Delta_j$ in Section
\ref{subsec.Delta}. The essential idea is contained in Proposition
\ref{prop-repr}:
\begin{eqnarray*}
& &a_{n,p;j}(\sigma)\hat{\beta}_{\mathrm{corr};j}
\\
&&\quad= a_{n,p;j}(\sigma) (P_{\bx})_{jj} -
a_{n,p;j}(\sigma) \sum_{k\neq j}
(P_{\bx})_{jk}\bigl(\hat{\beta}_{\mathrm{init};k} -
\beta^0_k\bigr) + a_{n,p;j}(\sigma)
Z_j + \mbox{negligible term}.
\end{eqnarray*}
There are three cases. If
%
\begin{equation}
\label{case1} a_{n,p;j}(\sigma) \sum_{k\neq j}
(P_{\bx})_{jk}\bigl(\hat{\beta}_{\mathrm{init};k} -
\beta^0_k\bigr) = o_P(1),
\end{equation}
a correction with the bound $\Delta_j$ would not be necessary, but of
course, it does not hurt in terms of type I error control.
If
%
\begin{equation}
\label{case2} a_{n,p;j}(\sigma) \sum_{k\neq j}
(P_{\bx})_{jk}\bigl(\hat{\beta}_{\mathrm{init};k} -
\beta^0_k\bigr) \asymp V,
\end{equation}
for some non-degenerate random variable $V$, the correction with the bound
$\Delta_j$ is necessary and assuming that $\Delta_j$ is of the same order
of magnitude as $V$, we have a balance
between $\Delta_j$ and the stochastic term $a_{n,p;j}(\sigma) Z_j$.
In the last
case where
%
\begin{equation}
\label{case3} a_{n,p;j}(\sigma) \sum_{k\neq j}
(P_{\bx})_{jk}\bigl(\hat{\beta}_{\mathrm{init};k} -
\beta^0_k\bigr) \to \infty,
\end{equation}
the bound $\Delta_j$ would be the dominating element in the $p$-value
construction. We show in Figure~\ref{fig-supp1} that there is empirical
evidence that (\ref{case2}) applies most often.
%

\renewcommand{\thefigure}{\arabic{figure}}
\setcounter{figure}{2}
\begin{figure}

\includegraphics{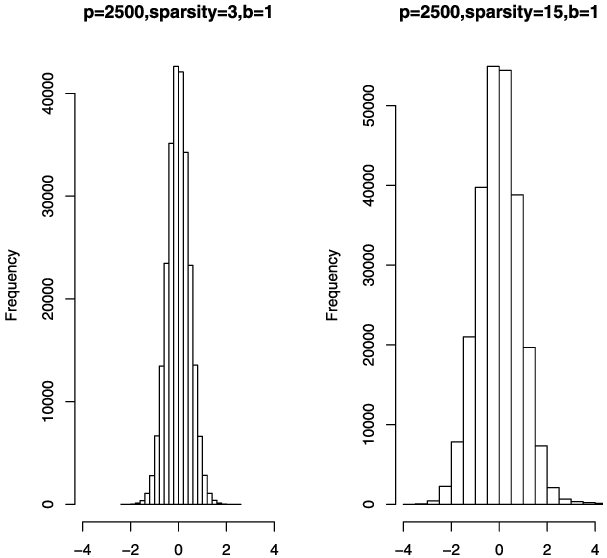}

\caption{Histogram of projection bias $a_{n,p;j}(\sigma)\sum_{k\neq j}
(P_{\bx})_{jk}(\hat{\beta}_{\mathrm{init};k} - \beta^0_k)$ over
all values
$j=1,\ldots,p$ and over 100 independent simulation runs. Left: model
(M2), $p=2500$, $s=3$, $b=1$; Right: model (M2), $p=2500$, $s=15$,
$b=1$.}\label{fig-supp1}
\end{figure}

Case (\ref{case3}) is comparable to a crude procedure which makes a hard
decision about relevance of the underlying coefficients:
\begin{eqnarray*}
\mbox{if } a_{n,p;j}(\sigma) |\hat{\beta}_{\mathrm{corr};j}| >
\Delta_j \mbox{ holds, then $H_{0,j}$ is rejected},
\end{eqnarray*}
and the rejection would be ``certain'' corresponding to a $p$-value with
value equal to $0$; and in case of a ``$\le$'' relation, the corresponding
$p$-value would be set to one.
This is an analogue to the thresholding rule:
%
\begin{eqnarray}
\label{hard-rule} \mbox{if}\ |\hat{\beta}_{\mathrm{init};j}| >
\Delta_{\mathrm
{init}}\ \mbox{holds, then $H_{0,j}$ is rejected},
\end{eqnarray}
where $\Delta_{\mathrm{init}} \ge\|\hat{\beta}_{\mathrm{init}} -
\beta^0\|_{\infty}$, e.g. using a bound where $\Delta_{\mathrm
{init}} \ge
\|\hat{\beta}_{\mathrm{init}} - \beta^0\|_{1}$. For example,
(\ref{hard-rule}) could be the variable selection estimator with the
thresholded
Lasso procedure \citep{geer11}. An accurate construction of
$\Delta_{\mathrm{init}}$ for practical use is almost impossible: it
depends on $\sigma$ and in a complicated way on the nature of the design
through e.g. the compatibility constant, see (\ref{oracle-ineq}).

Our proposed bound $\Delta_j$ in (\ref{bound1}) is very simple. In principle,
its justification also depends on a bound for $\|\hat{\beta}_{\mathrm
{init}} -
\beta^0\|_{1}$, but with the advantage of ``robustness''. First, the bound
$a_{n,p;j}(\sigma) \max_{k \neq j}|(P_{\bx})_{jk}| \|\hat{\beta
}_{\mathrm{init}} -
\beta^0\|_1$ appearing in (\ref{crudebound}) is not depending on
$\sigma$
anymore (since $\|\hat{\beta}_{\mathrm{init}} -
\beta^0\|_1$ scales linearly with $\sigma$). Secondly, the inequality in
(\ref{crudebound}) is crude implying that $\Delta_j$ in (\ref
{bound1}) may
still satisfy assumption \textup{(A)} even if the bound of
$\|\hat{\beta}_{\mathrm{init}} - \beta^0\|_1$ is misspecified and too
small. The
construction of $p$-values as in (\ref{pvalue1}) and (\ref{pvalue2}) is much
better for practical purposes (and for simulated examples) than using a rule
as in (\ref{hard-rule}).

\end{appendix}

\section*{Acknowledgements}
I would like to thank Cun-Hui Zhang for fruitful discussions and Stephanie
Zhang for providing an R-program for the Scaled Lasso.

%


\printhistory

\end{document}